\documentclass[prd,aps, preprint,10pt,nofootinbib]{revtex4-1}
\pdfoutput=1
\usepackage{geometry} 
\usepackage{graphicx}
\usepackage{epstopdf}
\usepackage{verbatim}
\usepackage{color}
\usepackage{slashed}
\usepackage{bbold}
\usepackage[title]{appendix}
\usepackage{subfig}
\usepackage {amssymb,amsfonts,amsmath,mathtools}
\usepackage[colorlinks, citecolor=green, linkcolor=blue]{hyperref}

\newcommand{\BigO}[1]{\ensuremath{\operatorname{\mathcal{O}}\left(#1\right)}}

\begin{document} 

\title{Domain walls and gravitational waves in the Standard Model}

\author{Tomasz~Krajewski}
\email{Tomasz.Krajewski@fuw.edu.pl}
\author{Zygmunt~Lalak}
\email{Zygmunt.Lalak@fuw.edu.pl}
\author{Marek~Lewicki}
\email{Marek.Lewicki@fuw.edu.pl}
\author{Pawe\l~Olszewski}
\email{Pawel.Olszewski@fuw.edu.pl}
\affiliation{Institute of Theoretical Physics, Faculty of Physics, University of Warsaw, ul. Pasteura 5, Warsaw, Poland\vspace{20.pt}}

\begin{abstract}
We study domain walls which can be created in the Standard Model under the assumption that it is valid up to very high energy scales.
We focus on domain walls interpolating between the physical electroweak vacuum and the global minimum appearing at very high field strengths. The creation of the network which ends up in the electroweak vacuum percolating through the Universe is not as difficult to obtain as one may expect, although it requires certain tuning of initial conditions.
Our numerical simulations confirm that such domain walls would swiftly decay and thus cannot dominate the Universe.
We discuss the possibility of detection of gravitational waves produced in this scenario. We have found that for the standard cosmology the energy density of these gravitational waves is too small to be observed in present and planned detectors.
\end{abstract}

\maketitle

\section{Introduction}
After the discovery of the Higgs boson at the Large Hadron Collider \cite{Aad:2012tfa,Chatrchyan:2012ufa} all of the parameters of the Standard Model (SM) are known with high accuracy.
This knowledge has enabled the quantitative study of the RG improved effective SM potential, which revealed an interesting structure at field strengths higher than approximately $10^{11}$ GeV where the potential drops below the electroweak vacuum values and leads to a new minimum at superplanckian field strengths.
The possibility of the tunnelling from a~physical electroweak symmetry breaking minimum (EWSB) to the deeper minimum located at superplanckian values of the Higgs field and corresponding instability of the observed vacuum have been widely discussed in the literature \cite{Lalak:2014qua,Buttazzo:2013uya,Degrassi:2012ry,Ellis:2009tp,Espinosa:2007qp,Casas:2000mn,Casas:1996aq,Casas:1994qy,Sher:1988mj}. 

It is possible that in the early Universe, for example during inflation \cite{Hook:2014uia,Kobakhidze:2013tn}, the Higgs field fluctuations were large enough to overcome the potential barrier between the two minima. This would result in creation of a~network of domain walls which interpolate between areas of the Universe occupied by the Higgs field laying in different minima.
Evolution of these structures is our main interest. To remain independent from a specific model of the early Universe we investigate dependence of our results on initial conditions: a~mean value of the field strength and an~amplitude of fluctuations.

If minima of the potential are not degenerate, domain walls interpolating between these minima are unstable and they annihilate on a~time scale which depends on the fraction of the space occupied by the field strength corresponding to the global minimum of the potential, the bias between minima (i.e. the difference between values of the potential at minima) and the value of the derivative on both sides of the local maximum separating the minima \cite{Lalak:2007rs}. 
Numerical simulations based on Press, Ryden, Spergel (PRS) algorithm \cite{Press:1989yh} revealed that the conformal time needed for networks of the SM domain walls to decay is relatively short and of the order of $10^{-9}\ \frac{\hbar}{\textrm{GeV}}$.

The effective equation of state of a~network of cosmological domain walls is generically predicted with coefficient $-2/3 < w_X < -1/3$. The energy density of the network of stable domain walls decreases (with the expansion) slower then the energy density of both radiation and dust, acting as dark energy, so long lived domain walls tend to dominate the Universe. However dark energy with such equation of state is ruled out by the present data. Moreover domain walls which pose a significant fraction of the energy density of the Universe at recombination would produce unacceptably large fluctuations of the Cosmic Microwave Background Radiation (CMBR). However, very short lifetime of the SM domain walls makes them consistent with the observational data.

During the process of the decay of domain walls the energy of the field is transferred to other degrees of freedom, both SM particles and gravitational waves (GWs) can be produced. The recent observation of GWs at the LIGO and the Virgo experiments \cite{Abbott:2016blz} promoted spectrum of GWs to one of the most promising observables which could in principle probe domain walls in the early Universe. Moreover GWs produced from networks of domain walls could partially polarise CMBR, marking it with a distinctive pattern.
We estimate the spectrum of GWs produced during the decay of SM domain walls using two methods: a~semi-analytical approximation used in previous studies \cite{Kitajima:2015nla,Hiramatsu:2013qaa} and a~direct calculation utilising a lattice simulation. 
Both approaches predict the energy density of GWs orders of magnitude smaller than the sensitivity of present and planned detectors of GWs. Our result for the case of the standard cosmology excludes the possibility of detection of GWs produced by domain walls of the Higgs field in near future. The present energy density of GWs produced by domain walls could be greater if the evolution of the Universe before formation of domain walls was different then in the standard scenario. However this possibility is hard to investigate using lattice simulations due to their small dynamical range and require semi-analytical extrapolation.

The paper is organised as follows. In section \ref{SM_potential} we briefly remind the form of the RG improved effective potential in the SM. The method we use to estimate the width of domain walls is derived in subsection \ref{toy_model} and is applied to the SM in subsection \ref{SM}. Section \ref{decay_time} is dedicated to the estimation of the lifetime of networks of domain walls. We discuss the dependence of the lifetime on initial conditions: the time of formation in subsection \ref{start_time}, the average value of the field strength and its standard deviation at the initialisation in subsection \ref{mean_value}. Section \ref{algorithm} contains brief derivation of the numerical algorithm used to calculate the spectrum of GWs emitted during the decay of domain walls. This spectrum is presented in section \ref{spectrum}.

\section{RG improved SM effective potential\label{SM_potential}}
Previous studies of the evolution of domain walls focused on the dynamics driven by a~classical action $S_{cl}$. However quantum corrections may introduce significant modifications for large enough field strengths, as it happens in the case of the SM where a~new minimum of the effective potential is generated at superplanckian field strengths. The starting point for our work is observation that the expectation value of the field strength $\Psi_{cl}(x)$ satisfies the following equation of motion (eom):\footnote{We consider the space-time dependent expectation value $\Psi_{cl}$.}
\begin{equation}
\frac{\delta}{\delta \Psi_{cl}(x)} \Gamma_{\text{eff}}[\Psi_{cl}]=0, \label{eom_general}
\end{equation}
where $\Gamma_{\text{eff}}$ is the 1PI effective action. The effective potential $V_{\text{eff}}$ is determined by the value of $\Gamma_{\text{eff}}$ at the expectation value which is constant in time and space $\Psi_{cl}(x)=\psi_{cl}$ by the equation:
\begin{equation}
V_{\text{eff}}(\psi_{cl})= \lim_{V \to \infty} \lim_{T \to \infty} -\frac{\Gamma_{\text{eff}}[\psi_{cl}]}{V T},
\end{equation}
where $V$, $T$ are respectively the volume and the time extent of the compact region in which we consider our theory.\footnote{The presented limit is the thermodynamical limit of the theory. The $V_{\text{eff}}$ is intensive parameter and $\Gamma_{\text{eff}}$ is extensive, so the former tends to the constant and the later diverges.} The constant field $\psi_{cl}$ with strength equal to the extremum of the effective potential $V_{\text{eff}}$ is the solution of eom from eq. \eqref{eom_general}. 
However constant fields are not the only solutions of eq. \eqref{eom_general} and space-time depended solutions are our main interest. 

The 1PI effective action $\Gamma_{\text{eff}}$ is rarely known exactly and in practical applications must be approximated. We have used the following approximation:
\begin{equation}
\Gamma_{\text{eff}}[\phi] \approx \int d^4x \sqrt{|\det g|} \left[ \frac{1}{2} g^{\mu \nu} \partial_\mu \phi \partial_\nu \phi - V_{\text{SM}}(\phi) \right], \label{action_spproximation}
\end{equation}
where $\phi$ is a~real scalar field which models the Higgs field in our simulations and $V_{\text{SM}}$ is the effective potential of the SM in zero temperature. The eom for the approximated effective action proposed in eq. \eqref{action_spproximation} is of the form:
\begin{equation}
\frac{\partial^2 \phi}{\partial \eta^2} + \frac{2}{a} \left(\frac{d a}{d \eta}\right) \frac{\partial \phi}{\partial \eta} - \Delta \phi + a^2 \frac{\partial V_{\text{eff}}}{\partial \phi}=0, \label{SM_eom}
\end{equation}
assuming the Friedman-Robertson-Walker metric background:
\begin{equation}
g = dt^2 - a^2(t) \delta_{ij} dx^i dx^j = a^2(\eta) \left(d\eta^2 -\delta_{ij} dx^i dx^j\right), 
\end{equation}
where Latin indices correspond to spatial coordinates, $t$ is cosmic time and $\eta$ denotes conformal time (such that $d \eta = \frac{1}{a(t)} dt$). For the purpose of this paper we assumed that the time evolution of the scale factor $a(\eta)$ is as in the radiation domination epoch i.e. grows linearly:
\begin{gather}
\frac{d a}{d \eta}= a(\eta)^2 H(\eta) = const =: \dot{a}_{in},\\
a(\eta) = \dot{a}_{in} \eta + a_{in}, \label{scale_factor}
\end{gather}
where $H(\eta)$ is the value of the Hubble constant at the conformal time $\eta$ and $a_{in}$ is a~constant which sets the value of the scale factor $a$ at the time when radiation starts to dominate the Universe. The back reaction from domain walls to the metric tensor was neglected. This assumption is supported by the following reasoning. The present experimental data shows that the Universe is nearly flat, so it was nearly flat in the past too. That means that a~total energy density was nearly equal to the critical density
\begin{equation}
\rho_c=3 H^2 M_{Pl}^2,
\end{equation}
where $H$ denote the Hubble constant. Comparing $\rho_c$ with the difference between values of $V_{\text{SM}}$ at the maximum $v_{max}$ and at the EWSB minimum $v_{EWSB}$ which is of the order of $V_{\text{SM}}(v_{max})- V_{\text{SM}}(v_{EWSB}) \sim 10^{37} \textrm{GeV}^4$ (with one sigma confidence region ranging from $10^{34}\textrm{GeV}^4$--$10^{42}\textrm{GeV}^4$) one finds that the Universe had the total density of the same order when the Hubble constant was of the order of $1 \frac{\textrm{GeV}}{\hbar}$ which is much smaller then the Hubble constant at the end of our simulation \eqref{simulation}. The spatial fluctuations of the energy density induced by Higgs domain walls constituted a~very small fraction of the total energy density.

The precise shape of the effective potential $V_{\text{SM}}$ has been long and extensively studied, see e.g. \cite{Sher:1988mj} and \cite{Casas:1996aq}. Since the discovery of the Higgs boson all of parameters of the SM potential are known accurately enough to enable quantitative study of $V_{\text{SM}}$. As an approximation of $V_{\text{SM}}$ we have used the following expression:
\begin{gather}
V_{\text{SM}}(h;\mu) = \frac{1}{2} m^2(\mu) h^2 + \frac{1}{4} \lambda_{\text{eff}}(h; \mu) h^4 \\
\lambda_{\text{eff}} = \lambda(\mu) + \lambda_{\text{eff}}^{(1)}(h;\mu) + \lambda_{\text{eff}}^{(2)}(h;\mu),\label{eqn:pot}
\end{gather}
where $\lambda_{\text{eff}}^{(n)}(h;\mu)$ denotes $n$-loop correction to the quartic term, $h$ is the Higgs field strength. It depends on $\mu$ through the running couplings $\lambda$, $g_1$, $g_2$, $g_3$, $y_{top}$. We neglect very small Yukawa couplings of leptons and quarks, keeping only the large top Yukawa coupling $y_{top}$.

The field strength $h$ appears in $\lambda_{\text{eff}}$ only via the logarithmic corrections $\log(h/\mu)$. We use renormalization group equations (RGEs) to improve validity of the loop expansion and we substitute $\mu = \langle |h| \rangle$ to ensure the corrections are small. 

\begin{equation}
\widetilde{V}_{\text{SM}} (h) \equiv V_{\text{SM}} (h, |h|).
\end{equation}

The loop level up to which we work is two for $\lambda_{\text{eff}}$ and three for RGEs. We have used expressions for $\lambda_{\text{eff}}^{(1)}$, $\lambda_{\text{eff}}^{(2)}$, RGEs of $m$, $\lambda$, $g_1$, $g_2$, $g_3$, $y_{top}$ and their values renormalised at the scale $\mu=m_{top}$ taken from \cite{Buttazzo:2013uya}.

\begin{figure}[t]
\includegraphics[width=0.49\textwidth]{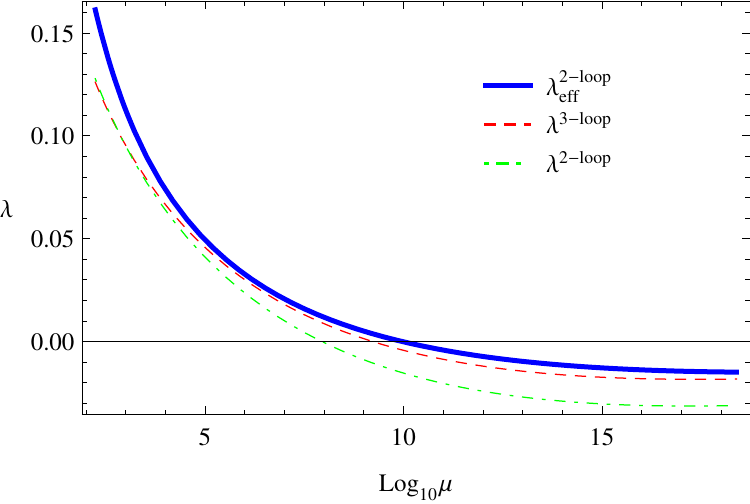} 
\includegraphics[width=0.49\textwidth]{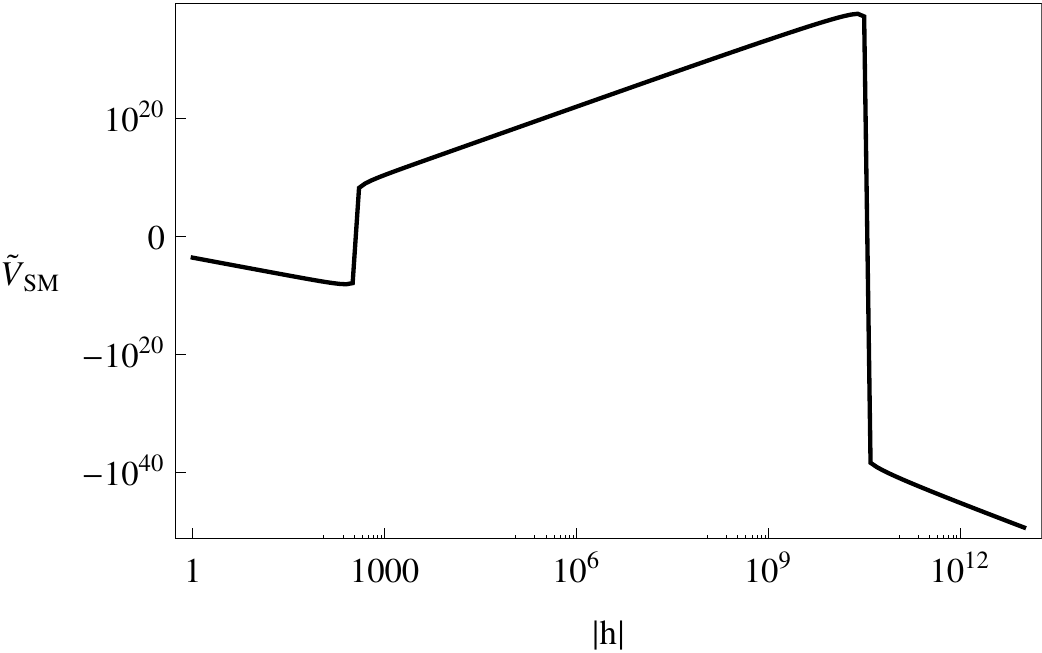} 
\caption{Running coupling $\lambda$ together with the effective coupling from eq. \eqref{eqn:pot} (left panel) and the resulting SM effective potential $\widetilde{V}_{\text{SM}}$ (right panel).
\label{VSM_potential} }
\end{figure}

The potential $\widetilde{V}_{\text{SM}}$ used in our simulations is presented in figure \ref{VSM_potential}. It resembles a mexican hat only at scales $\mathcal{O}(10^2\ \textrm{GeV})$. At larger scales it is determined by the behaviour of $\lambda_{\text{eff}}$, which decreases all the way from $\lambda_{\text{eff}}(\sim 10^2\ \textrm{GeV}) \sim 0.1$, down to $\lambda_{\text{eff}}(\sim 10^{17}\ \textrm{GeV}) \sim -0.02$. Along that way the potential reaches a maximum around the value $|h|_\text{max} \sim 10^{10}\ \textrm{GeV}$, $V_\text{max} \sim 10^{34}\ \textrm{GeV}^4$. Slope of the potential is much steeper for field values $|h|>|h|_\text{max}$ than it is for $|h|<|h|_\text{max}$, and potential quickly decreases below value at the electroweak (EW) minimum for field strengths $|h| \gtrsim 10^{10}\textrm{GeV}$. 

Assuming validity of the SM up to such high scales, one witnesses an intricate vacuum structure, which in principle allows for formation and non-trivial evolution of domain walls in the early Universe. At the same time the need to control dynamics of the field throughout disparate energy scales poses a considerable technical challenge.
Minima of the SM effective potential lay at very different scales the EW minimum is at the field strength of the order of $246\ \textrm{GeV}$, second minimum is positioned at superplanckian values of field strength and the local maximum separating these minima is at the field strength of the order of $10^{10}\ \textrm{GeV}$. Moreover minima of the SM potential are highly non-degenerate. Previous studies of the evolution of domain walls have concentrated on the case of nearly degenerate minima.

The RG improved effective potential of the SM is known only as a~numerical solution of the RGEs and had to be modelled numerically in our simulations. Due to experimental uncertainties of parameters of the SM the effective potential is known only with certain precision. The main impact on the inaccuracy of the SM potential comes from uncertainties $\sigma_{M_t}$, $\sigma_{M_h}$ in determination of masses of the top quark $M_t$ and the Higgs boson $M_h$. We have adopted a~conservative estimation $\sigma_{\partial V}$ of this inaccuracy of the form:
\begin{multline}
\sigma_{\partial V} (|h|) := \max\left(\left|\frac{\partial V_{\text{SM}}^{M_t+\sigma_{M_t},M_h}}{\partial |h|}(|h|) - \frac{\partial V_{\text{SM}}^{M_t-\sigma_{M_t},M_h}}{\partial |h|}(|h|)\right|,\right.\\
\left.\left|\frac{\partial V_{\text{SM}}^{M_t,M_h+\sigma_{M_h}}}{\partial |h|}(|h|)-\frac{\partial V_{\text{SM}}^{M_t,M_h-\sigma_{M_h}}}{\partial |h|}(|h|)\right|\right),
\end{multline}
where $\frac{\partial V_{\text{SM}}^{M_1,M_2}}{\partial |h|}(|h|)$ denotes the value of derivative of the RG improved potential at point $|h|$ calculated with mass of the top quark equal to $M_1$ and mass of the Higgs boson $M_2$. 

For the measure $\sigma_{num}$ of uncertainty of our numerical model $\frac{\partial V}{\partial |h|}_{num}$ of $\frac{\partial V_{\text{SM}}}{\partial |h|}$ we adopted the following expression:
\begin{equation}
\sigma_{num}\left[\frac{\partial V}{\partial |h|}_{num}\right]=\sup_{|h|\in [1\ \textrm{GeV}, 2 M_{Pl}]} \frac{\left|\frac{\partial V}{\partial |h|}_{num}(|h|)-\frac{\partial V_{\text{SM}}}{\partial |h|}(|h|)\right|}{\sigma_{\partial V} (|h|)}.
\end{equation}
Our numerical model gives the value of $\sigma_{num}$ less than $7\%$.
\section{The width of domain walls\label{width}}
\subsection{Toy model\label{toy_model}}
Before proceeding to calculations within the SM we will describe our general setup in a~simple toy model, 
\begin{equation}
\mathcal{L}=\frac{1}{2} \partial_\mu \phi \partial^\mu \phi - V_0 \left(\frac{\phi^2}{{\phi_0}^2}-1\right)^2 -\epsilon V_0 \phi. \label{toy_lagrangian_density}
\end{equation}
The above Lagrangian density \eqref{toy_lagrangian_density} with $\epsilon = 0$ exhibits a $\mathbb{Z}_2$ symmetry realised by $\phi \mapsto -\phi$. 
For $\epsilon = 0$ two minima are degenerated, while non-zero $\epsilon$ breaks this degeneracy and renders one of them unstable.

The eom derived from \eqref{toy_lagrangian_density} in the Minkowski gravitational background takes the form:
\begin{equation}
\frac{\partial^2 \phi}{{\partial t}^2} -\Delta \phi = - \frac{\partial V}{\partial \phi} = 4 V_0 \left(\frac{\phi^2}{{\phi_0}^2}-1\right)\frac{\phi}{{\phi_0}^2} -\epsilon V_0. \label{toy_eom}
\end{equation}

We are interested in a time independent solution (soliton solution). We will consider planar walls i.e. solutions with a~translational symmetry in two space dimensions. Assuming \mbox{$\phi(t,x,y,z)=\varphi(x)$}, our Lagrangian density \eqref{toy_lagrangian_density} (with $\epsilon=0$) simplifies to
\begin{equation}
\mathcal{L}=-\frac{1}{2} \varphi'^2 - V\left(\varphi\right)=-\frac{1}{2} \varphi'^2- V_0 \left(\frac{\varphi^2}{{\phi_0}^2}-1\right)^2, \label{toy_lagrangian_density_simplified}
\end{equation}
where prime is a derivative with respect to $x$.
This Lagrangian density has a translational symmetry in $x$ and the corresponding conservation law. The associated conserved quantity is
\begin{equation}
E=\frac{1}{2} \varphi'^2 - V\left(\varphi\right)=\frac{1}{2} \varphi'^2- V_0 \left(\frac{\varphi^2}{{\phi_0}^2}-1\right)^2. \label{energy}
\end{equation}
Using conservation of $E$ we get first-order differential equation:
\begin{equation}
\varphi' = \pm \sqrt{2\left(E+V\left(\varphi\right)\right)}=\pm \sqrt{2\left(E+V_0 \left(\frac{\varphi^2}{{\phi_0}^2}-1\right)^2\right)},
\end{equation}
which can be easily integrated,
\begin{equation}
x(\varphi_2)-x(\varphi_1)= \pm \int_{\varphi_1}^{\varphi_2} \frac{d\varphi}{\sqrt{2\left(E+V\left(\varphi\right)\right)}}= \pm \int_{\varphi_1}^{\varphi_2} \frac{d\varphi}{\sqrt{2\left(E+V_0 \left(\frac{\varphi^2}{{\phi_0}^2}-1\right)^2\right)}}, \label{integral}
\end{equation}
for appropriate values of $\phi_1$ and $\phi_2$. Choosing $x_1=\phi_1=0$ we get our soliton solution
\begin{equation}
\varphi(x) = \phi_0 \tanh \left(\frac{\sqrt{2 V_0}}{\phi_0}x\right) = \phi_0 \tanh \left(\frac{\pi x}{w_0}\right), \label{soliton}
\end{equation}
where $w_0=\frac{\pi \phi_0}{\sqrt{2 V_0}}$ is a~width of the wall.
We can also calculate surface potential energy, using
\begin{equation}
\sigma(x_1,x_2):=\int_{x_1}^{x_2} V(\varphi(x)) dx = \int_{\varphi(x_1)}^{\varphi(x_2)}\frac{V(\varphi) d\varphi}{\sqrt{2\left(E+V\left(\varphi\right)\right)}}. \label{tension}
\end{equation}
Most of the energy of the wall is concentrated at distances of the order of $w_0$ from the center of the wall,
\begin{equation}
\frac{\sigma(-\frac{w_0}{2},\frac{w_0}{2})}{\sigma(-\infty,\infty)} \approx 97 \%.
\end{equation}
This justifies the estimation of the domain wall thickness by the quantity $w_0$.

Finally it is important to stress that the time-independent soliton solution for a~planar domain wall does not exist if the minima of the potential are non-degenerate. We show this fact in appendix \ref{app:nosoliton}.

\subsection{Standard Model\label{SM}}
The estimation of the physical width of domain walls is critical for numerical simulations of their dynamics. The width must be at least a~few times larger than the lattice spacing (i.e. the physical distance between neighbouring points) used in the simulation in order to assure sufficient accuracy to model profiles of walls. On the other hand, if we choose the lattice spacing too small (walls will spread over too many lattice points) only few walls will fit into finite lattice. If only small number of walls will be present on the lattice, then dynamics of the network of domain walls will be reproduced poorly in the simulation. Many authors \cite{Press:1989yh,Coulson:1995uq,Lalak:1996db,Oliveira:2004he,Lalak:2007rs,Kawasaki:2011vv,Leite:2011sc,Hiramatsu:2013qaa} used simulations with the physical width of walls varying from 2 to 100 lattice spacing.

\begin{figure}[t]
\centering
\includegraphics[width= 0.7\textwidth]{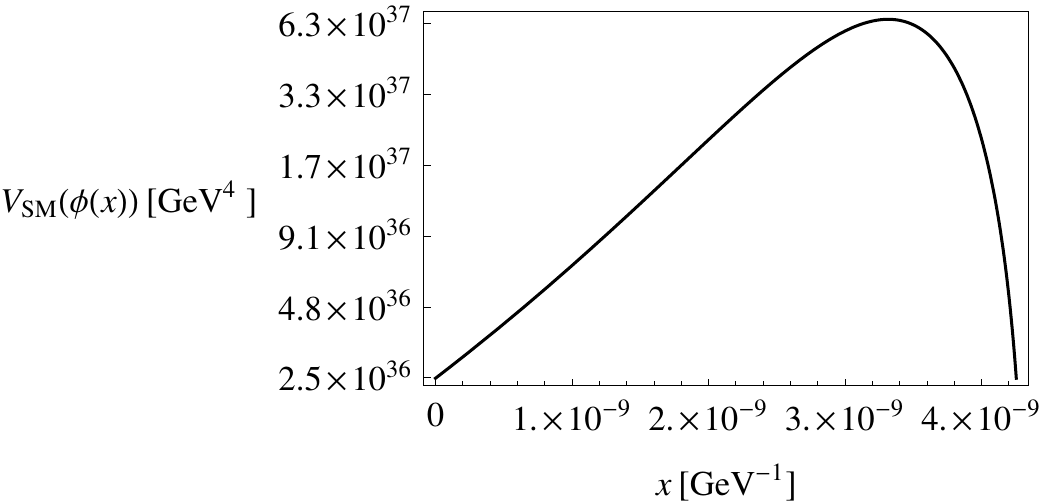}
\caption{The potential energy density $V_{\text{SM}}(\phi(x))$ of the planar SM domain wall as a~function of the distance $x$ in the direction perpendicular to the wall surface.\label{profile}}
\end{figure}

As discussed in section \ref{SM_potential} the Higgs effective potential has two non-equivalent minima with very different values of the potential. The analysis in appendix \ref{app:nosoliton} shows that in such a case the time-independent soliton solution of the eom does not exist. We estimate the width of domain walls in the SM using the approach presented in subsection \ref{toy_model} based on the first integral of the eom. Firstly we calculate the value $h_2$ of the Higgs field which gives the same value of the effective potential as the value $h_\textrm{EW}$ taken by the field in the electroweak vacuum and bigger than the local maximum. Next, we use an analogue of eq. \eqref{tension}:
\begin{equation}
\Sigma_\textrm{SM}(\varphi_1,\varphi_2):= \int_{\varphi_1}^{\varphi_2}\frac{V_{\textrm{SM}}(\varphi) d\varphi}{\sqrt{2\left(V_{\textrm{SM}}\left(\varphi\right)-V_{\textrm{SM}}(h_{\textrm{EW}})\right)}}, \label{SM_tension}
\end{equation}
and eq. \eqref{soliton}:
\begin{equation}
X(\varphi_2)-X(\varphi_1)= \int_{\varphi_1}^{\varphi_2} \frac{d\varphi}{\sqrt{2\left(V_{\textrm{SM}}\left(\varphi\right)-V_{\textrm{SM}}(h_{\textrm{EW}})\right)}}.\label{approx_solution}
\end{equation}
We have found values $\tilde{\varphi_1}$ and $\tilde{\varphi_2}$ such that $V_{\textrm{SM}}\left(\tilde{\varphi_1}\right)=V_{\textrm{SM}}\left(\tilde{\varphi_2}\right)$ and
\begin{equation}
\frac{\Sigma_\textrm{SM}(\tilde{\varphi_1},\tilde{\varphi_2})}{\Sigma_\textrm{SM}(h_\textrm{EW},h_2)}\approx 97 \%.
\end{equation}
Our estimation of the width of domain walls in the SM gives:\footnote{We are using the convention $\hbar=1=c$, so the lengths and times are expressed in units of $\textrm{GeV}^{-1}$.}
\begin{equation}
w_\textrm{SM}:=X(\tilde{\varphi_2})-X(\tilde{\varphi_1}) \approx 4 \times 10^{-9}\ \textrm{GeV}^{-1}. \label{SM_width}
\end{equation}
The potential energy density as a~function of the distance in the direction perpendicular to the domain wall surface for the approximated solution \eqref{approx_solution} is presented in figure \ref{profile}. The value $x=0$ on the horizontal axis corresponds to $\varphi(0)=\tilde{\varphi_1}$.

We chose the physical lattice spacing $l$ (i.e. the physical distance between neighbouring lattice sites) to be equal to $10^{-10}\ \textrm{GeV}^{-1}$, so the width of the walls is of the order of $40\ l$. We used consistently the units of $10^{10}$ GeV through the paper. 

\section{The Decay of domain walls\label{decay_time}}
Cosmological domain walls are subject to many experimental constraints. Generally the energy density of a network of domain walls is predicted to decreases slower then the energy density of both: the radiation and the dust, so long lived domain walls would dominate the Universe. The equation of state of the network of domain walls is restricted to $-2/3 < w_X < -1/3$, which is ruled out by the present data for a~single component Dark Energy. Domain walls which lived long enough to be present during the recombination would produce unacceptably large fluctuations of CMBR. Moreover domain walls formed by the Higgs field are constrained by the following requirement: the final product of their decay must be the EW vacuum. For these reasons the time and the final state of the process of the decay of SM domain walls are the most obvious observables which can be compared with measurements. 

Our simulations were started with three initial conditions:
\begin{itemize}
\item initial conformal time $\eta_{start}$,
\item initial mean value of the field strength $\theta$,
\item initial standard deviation $\sigma$.
\end{itemize}
Initial conformal time $\eta_{start}$ is determined by the time at which domain walls are formed in the early Universe. However the initial time of the simulation must be earlier, in order to smooth out the initial numerical fluctuations by the field evolution. The time of the formation of a network of domain walls can be determined from the evolution of statistical quantities calculated in the simulation. Our simulations were run with the initial conformal time ranging from \mbox{$10^{-4}\ l = 10^{-14}\ \textrm{GeV}^{-1}$} to \mbox{$10\ l = 10^{-9}\ \textrm{GeV}^{-1}$}. The comparison of the simulated dynamics of domain walls formed at different times are presented in subsection \ref{start_time}. 

On the basis of discussion in \cite{Coulson:1995uq} we assumed that an~initial distribution of the field strength is given by the probability distribution
\begin{equation}
P(\phi) = \frac{1}{\sqrt{2 \pi} \sigma} e^{-\frac{\left(\phi -\theta\right)^2}{2 \sigma^2}}. \label{gauss_distribution}
\end{equation}
We studied the evolution of networks of domain walls initialised with different values of $\theta$ and $\sigma$ in order to accommodate variety of processes leading to the formation of the walls. According to \cite{Lalak:2007rs} the final state and the time of the decay of networks of domain walls depends on the fraction of the space occupied by the field strength corresponding to the global minimum of the potential, the bias between minima (i.e. the difference between values of the potential at minima) and the value of the derivative on both sides of the local maximum separating the minima. In the case of the SM Higgs the scalar potential strongly favors the global minimum. As a result final states of the evolution of initial configurations dominated by field strengths from the basin of attraction of the global minimum are determined to be the superplanckian minimum which is excluded by experiments. However a~fate of configurations dominated by field strengths from the basin of attraction of the EWSB minimum are not a priori known. Thus we concentrated on initial conditions which produces such configurations. We considered values of parameters $\theta$ and $\sigma$ of the probability distribution \eqref{gauss_distribution} such that
\begin{equation}
\int_{v_{max}}^{\infty}  P (\phi)d \phi < \frac{1}{2},
\end{equation}
in order to produce initial configurations dominated by the EWSB vacuum. The analysis of the dependence of the decay time on the mean value and the standard deviation is given in subsection \ref{mean_value}.

Initial conditions cannot be deduced from the dynamics of domain walls and must be derived from a~model of the evolution of the early Universe (for example an~inflationary model). Our results can be thought as a constraint on the space of models of the early Universe.

\subsection{Dependence on the initial time $\eta_{start}$\label{start_time}}

\begin{figure}[t]
\subfloat[]{\label{time_1.5_volume}
\includegraphics[width=0.5 \textwidth]{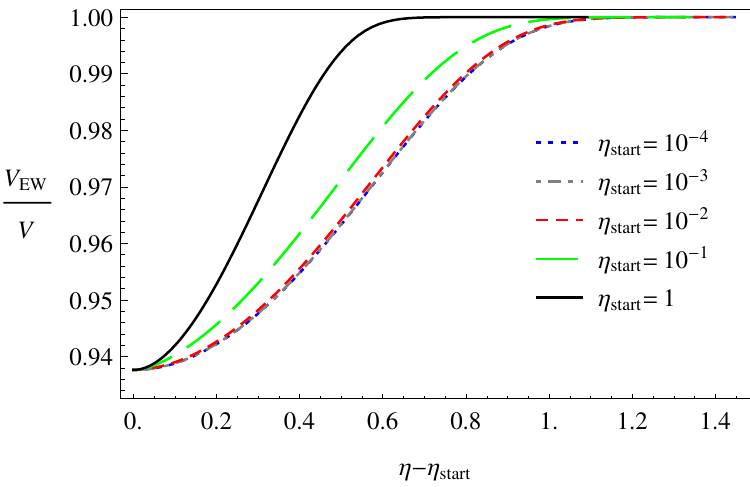}
 }
\subfloat[]{\label{time_3.0_volume}
\includegraphics[width=0.5 \textwidth]{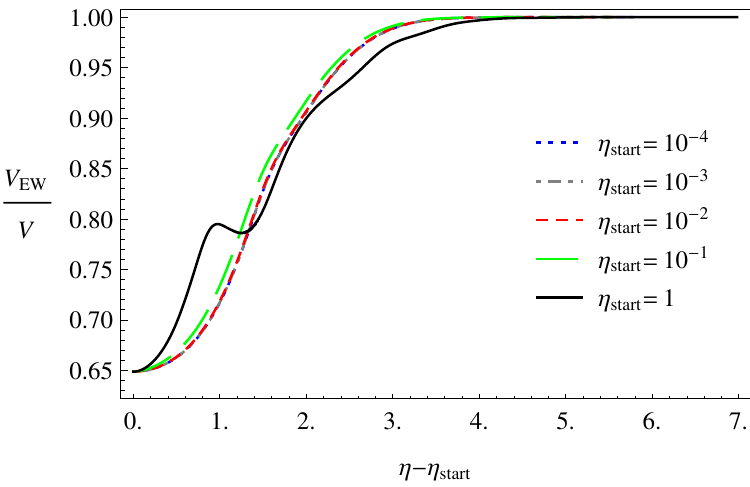}
 }
\caption{The time dependence of the fraction of lattice sites occupied by field on the electroweak side of the barrier $\frac{V_{EW}}{V}$ for small \mbox{\protect\subref{time_1.5_volume}} and moderate \mbox{\protect\subref{time_3.0_volume}} contribution of EW vacuum at initialisation.\protect\label{decay_time_small}}
\end{figure}

We run simulations for five different initialisation conformal times $\eta_{start}$ with $\theta=0$ and different $\sigma = \mathcal{O} \left(10^{10}\ \textrm{GeV}\right)$. Even though observed decay times are of the same order for all considered initialisation times, we observe a differences in the evolution of early formed (i.e. \mbox{$\eta_{start} < l = 10^{-10}\ \textrm{GeV}^{-1}$}) and late formed (i.e. $\eta_{start}>l$) domain walls. For initial conditions providing the configuration of the field where the electroweak vacuum strongly dominates, late domain walls decay faster then early ones. This issue is presented in figure \ref{time_1.5_volume} in which we plot the evolution of the ratio of the number of lattice sites occupied by the field strength belonging to the basin of attraction of the electroweak minimum to the size of the lattice $\frac{V_{EW}}{V}$. For more moderate contribution of EW vacuum in initial conditions the decay time of late domain walls can be longer than for early ones. The example of this situation is illustrated in figure \ref{time_3.0_volume}. 

\begin{figure}[t]
\subfloat[]{\label{time_3.0_surf}
\includegraphics[width=0.5 \textwidth]{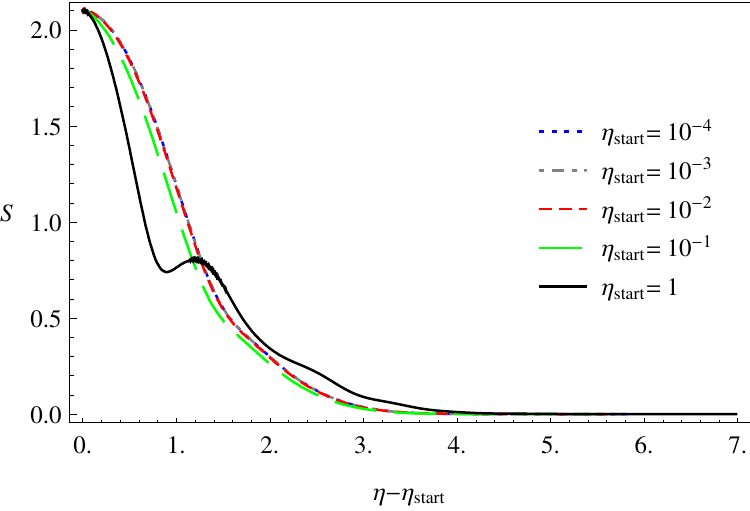}
 }
\subfloat[]{\label{time_3.0_ekin_vs_ewal}
\includegraphics[width=0.5 \textwidth]{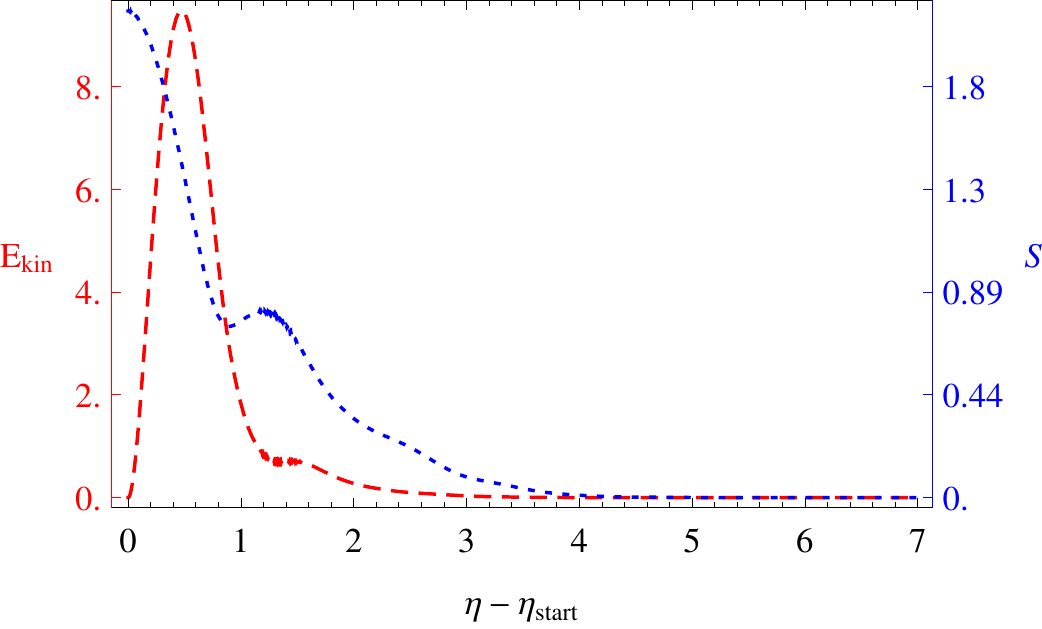}
 }
\caption{The time dependence of the surface area of domain walls per lattice site \mbox{\protect\subref{time_3.0_surf}} and comparison of the time dependence of kinetic energy of domain walls per lattice site (dashed) and surface area of domain walls per lattice site (dotted) for late initialisation time: $\eta_{start} = l$ \mbox{\protect\subref{time_3.0_ekin_vs_ewal}} for moderate contribution of the EW vacuum at the initialisation.\protect\label{decay_time_oscilations}}
\end{figure}

For late domain walls and the moderate contribution of the EW vacuum at the initialisation time $\eta_{start}$ we observe oscillations in the evolution of networks. The surface of domain walls increases and decreases several times before their complete decay. The example of such behaviour is given in figure \ref{time_3.0_surf} which presents the evolution of the surface in conformal time for the same simulations which were used in figure \ref{time_3.0_volume}. In figure \ref{time_3.0_ekin_vs_ewal} we plot kinetic (dashed) and potential (dotted) energies of field configurations as functions of conformal time $\eta$ for the late initialisation time $\eta_{start} = l$. The comparison of the time dependence of these two forms of the energy reveals that observed oscillations are due to the exchange between potential and kinetic energies of walls.

\begin{figure}[t]
\subfloat[]{\label{time_3.5_volume}
\includegraphics[width=0.5 \textwidth]{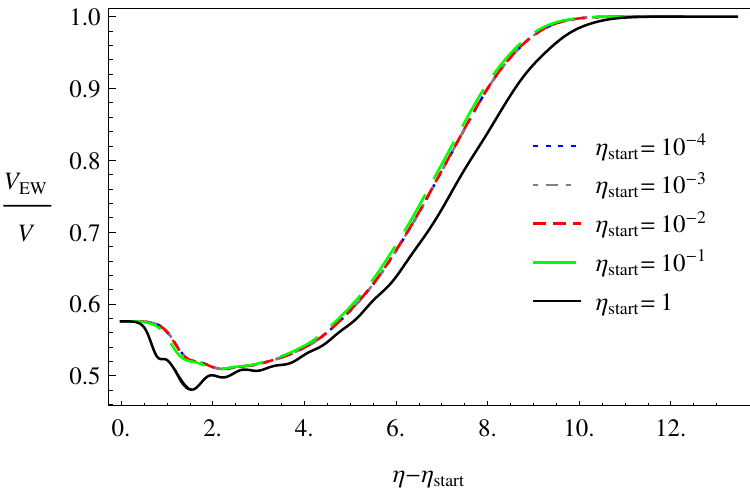}
 }
\subfloat[]{\label{time_4.0_volume}
\includegraphics[width=0.5 \textwidth]{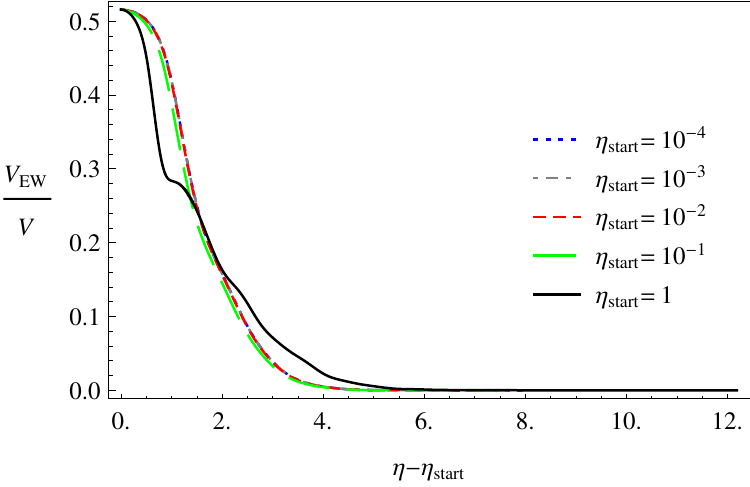}
 }
\caption{The time dependence of the fraction of lattice sites occupied by field on the electroweak side of the barrier $\frac{V_{EW}}{V}$ for nearly equal contributions of both vacua at initialisation time. \protect\label{decay_time_large}}
\end{figure}

For nearly equal contribution of both vacua at initialisation time late domain walls decay longer than early ones, as illustrated by the plot in figure \ref{time_3.5_volume}. It is worth stressing that for simulations presented in figure \ref{time_3.5_volume} $V_{EW}$ decreases initially, however finally the EW vacuum dominates the lattice. The decay of the domain walls leading to the final state without the EW vacuum is possible even for the initial configuration with slight dominance of the EW vacuum. These scenario is realised by the example shown in figure \ref{time_4.0_volume}.

We observe decay times of networks of domain walls ranging from \mbox{$0.7\ l = 7 \times 10^{-11}\ \textrm{GeV}^{-1}$} for initial conditions with the contribution of the high energy minimum of the order of a~few percents, up to \mbox{$1\ l = 1.3 \times 10^{-9}\ \textrm{GeV}^{-1}$} in the case of nearly equal contributions of both vacua at initialisation. In all cases the decay time displays the weak dependence on the initialisation time $\eta_{start}$.

\subsection{Dependence on mean value and variation of the field strength\label{mean_value}}
We considered cases when the sum of the mean field value $\theta$ and some fraction $\frac{1}{f}$ of the standard deviation $\sigma$ is equal to the position $v_{max}$ of the local maximum of the potential at the initialisation time $\eta_{start}$. We restricted the initial standard deviation to $10^{-3} \theta \le \sigma \le \theta$. 

\begin{figure}[t]
\subfloat[]{\label{mean_-4_volume_f=1}
\includegraphics[width=0.5 \textwidth]{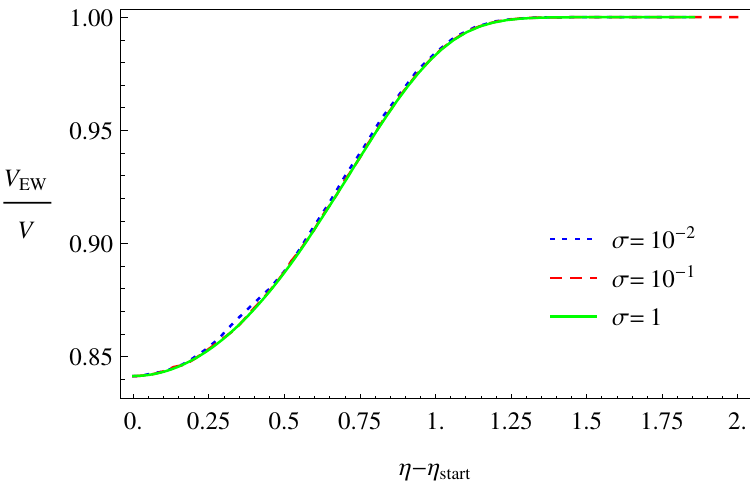}
 }
\subfloat[]{\label{mean_+1_volume_f=1}
\includegraphics[width=0.5 \textwidth]{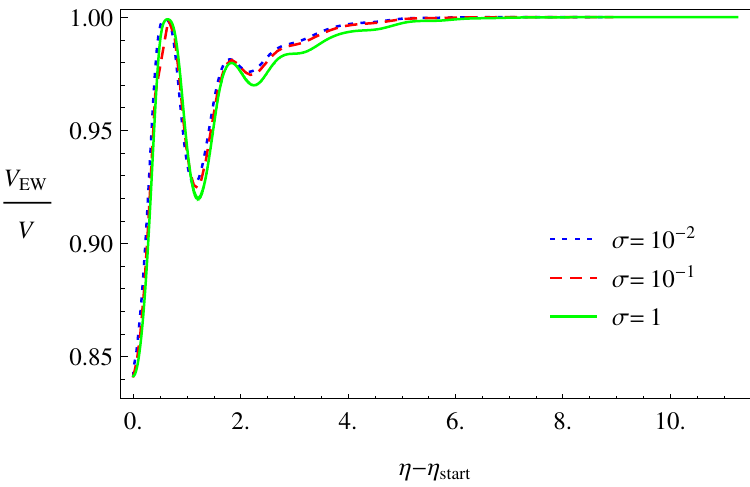}
 }
\caption{The time dependence of the fraction of lattice sites occupied by field on the electroweak side of the barrier $\frac{V_{EW}}{V}$ for different $\sigma$ for two initialisation times: \mbox{$\eta_{start}=10^{-4}\ l$} \mbox{\protect\subref{mean_-4_volume_f=1}} and \mbox{$\eta_{start}=10\ l$ \protect\subref{mean_+1_volume_f=1}}. \protect\label{mean_f=1}}
\end{figure}

For $f=1$ simulations start with ratio $\frac{V_{EW}}{V}$ equal to 84 \%. In this case the evolution of networks displays the weak dependence on the value of $\sigma$ and for all simulations the final state is the EW vacuum. Decay times in simulations with initialisation times \mbox{$10^{-4}\ l\le\eta_{start}\le 1\ l$} are all of the order of \mbox{$2\ l = 2 \times 10^{-10}\ \textrm{GeV}^{-1}$}. The time dependence of $\frac{V_{EW}}{V}$ for the network of domain walls initialised at the time \mbox{$\eta_{start}=10^{-4}\ l$} is presented in figure \ref{mean_-4_volume_f=1}. For late domain walls i.e. \mbox{$10\ l\le\eta_{start}\le 10^3\ l$}, similarly as for the case $\theta=0$ discussed in subsection \ref{start_time}, we observed oscillations. However differently than in the mentioned case there is discrepancy between decay times of early and late domain walls. Late domain walls decay in the time of the order of \mbox{$14\ l=1.4 \times 10^{-9}\ \textrm{GeV}^{-1}$}. The time dependence of $\frac{V_{EW}}{V}$ for the network of domain walls initialised at the time $\eta_{start}=10\ l$ is presented in figure \ref{mean_+1_volume_f=1}.

\begin{figure}[t]
\subfloat[]{\label{mean_-4_volume_f=2}
\includegraphics[width=0.5 \textwidth]{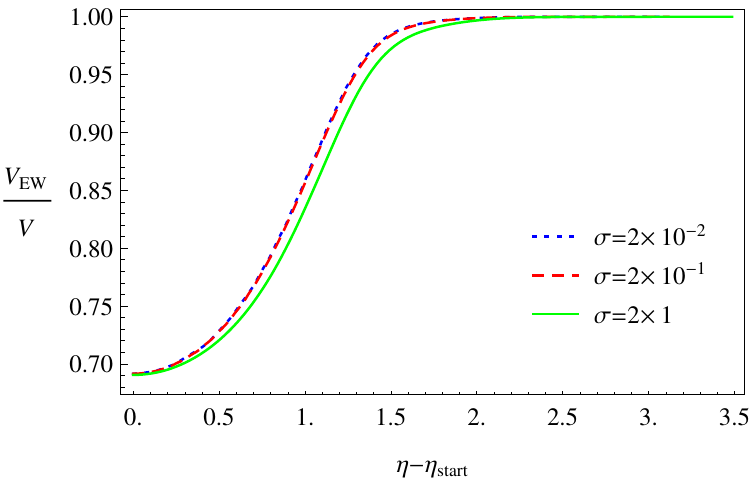}
 }
\subfloat[]{\label{mean_+1_volume_f=2}
\includegraphics[width=0.5 \textwidth]{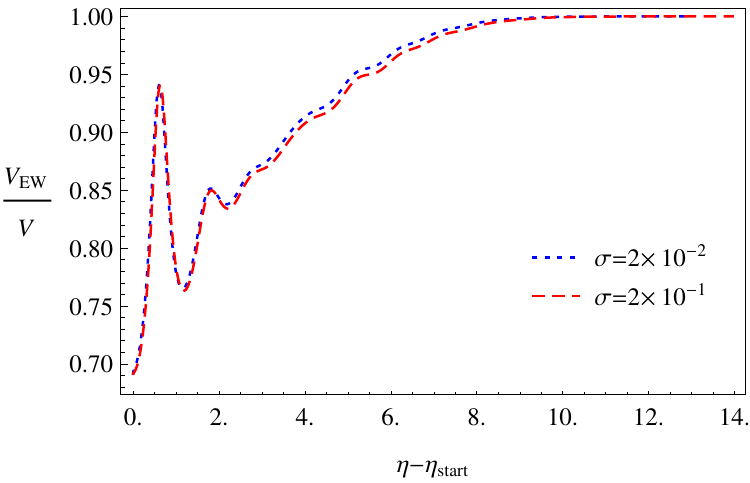}
 }
\caption{Time dependence of the fraction of lattice sites occupied by field on the electroweak side of the barrier $\frac{V_{EW}}{V}$ for different $\sigma$ for two initialisation times: $\eta_{start}=10^{-4}\ l$ \mbox{\protect\subref{mean_-4_volume_f=2}} and \mbox{$\eta_{start}=10\ l$ \protect\subref{mean_+1_volume_f=2}}. \protect\label{mean_f=2}}
\end{figure}

For $f=2$ simulations start with ratio $\frac{V_{EW}}{V}$ equal to 69 \%. As for the previous case the evolution of networks displays the weak dependence on the value of $\sigma$ and for all simulations the final state is the EW vacuum. Decay times for early domain walls are longer if $f=2$, then for $f=1$, and are of the order of $3.5\ l = 3.5 \times 10^{-10}\ \textrm{GeV}^{-1}$. For late domain walls decay times for both cases are of the order of $14\ l=1.4 \times 10^{-9}\ \textrm{GeV}^{-1}$. The time dependence of $\frac{V_{EW}}{V}$ for networks of domain walls initialised at the time $\eta_{start}=10^{-4}\ l$ is presented in figure \ref{mean_-4_volume_f=2} and for ones initialised at the time $\eta_{start}=10\ l$ is presented in figure \ref{mean_+1_volume_f=2}. For late domain walls i.e. $10\ l\le\eta_{start}\le 10^3\ l$ we observed oscillations.

\begin{figure}[t]
\subfloat[]{\label{mean_-4_volume_f=5}
\includegraphics[width=0.5 \textwidth]{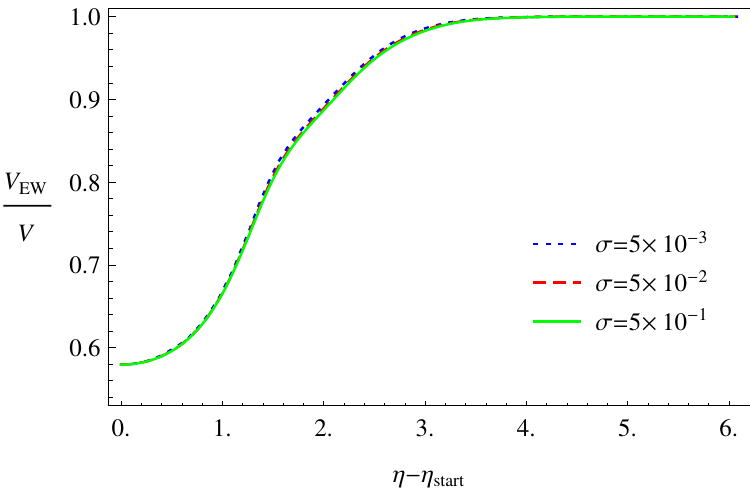}
 }
\subfloat[]{\label{mean_+1_volume_f=5}
\includegraphics[width=0.5 \textwidth]{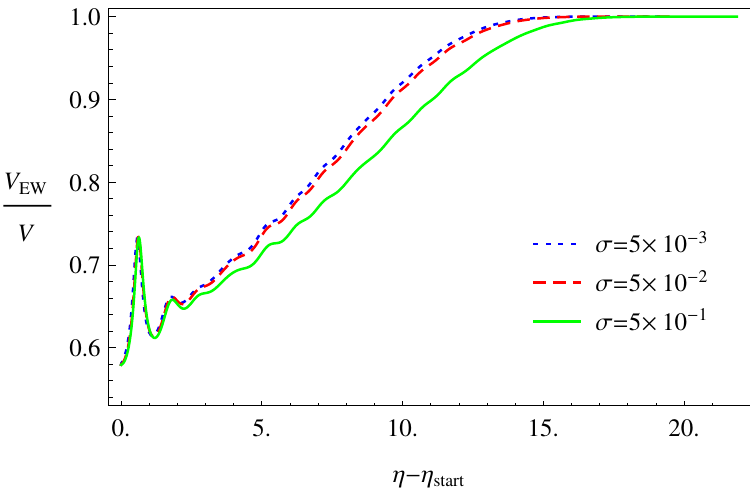}
 }
\caption{Time dependence of the fraction of lattice sites occupied by field on the electroweak side of the barrier $\frac{V_{EW}}{V}$ for different $\sigma$ for two initialisation times: $\eta_{start}=10^{-4}\ l$ \mbox{\protect\subref{mean_-4_volume_f=5}} and \mbox{$\eta_{start}=10\ l$ \protect\subref{mean_+1_volume_f=5}}. \protect\label{mean_f=5}}
\end{figure}

For $f=5$ simulations start with ratio $\frac{V_{EW}}{V}$ equal to 58 \%. For $f=5$ decay times for both early and late domain walls are longer then for $f=2$ and $f=1$. For early domain walls these times are of the order of $6\ l = 6 \times 10^{-10}\ \textrm{GeV}^{-1}$ and for late domain walls are of the order of $20\ l=2 \times 10^{-9}\ \textrm{GeV}^{-1}$. The time dependence of $\frac{V_{EW}}{V}$ for networks of domain walls initialised at the time $\eta_{start}=10^{-4}\ l$ is presented in figure \ref{mean_-4_volume_f=5} and for ones initialised at the time $\eta_{start}=10\ l$ is presented in figure \ref{mean_+1_volume_f=5}. For late domain walls i.e. $10\ l\le\eta_{start}\le 10^3\ l$ we observe oscillations.

\subsection{Dependence on the size of the lattice and statistical fluctuations}
\begin{figure}[t]
\centering
\begin{minipage}[t]{0.45\textwidth}
\includegraphics[width=\textwidth]{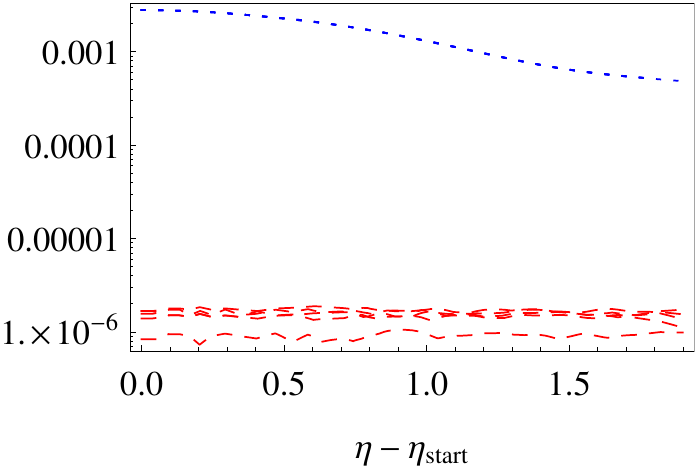}
\caption{Differences $\frac{|\bar{\phi}_i - \left\langle \bar{\phi} \right\rangle|}{\left\langle \bar{\phi} \right\rangle}$ (dashed) and standard deviations $\sigma_i$ (dotted) for five different runs on the lattice of the size $256^3$. \protect\label{statistical_error}}
\end{minipage}%
\hspace{0.5cm}
\begin{minipage}[t]{0.45\textwidth}
\includegraphics[width=\textwidth]{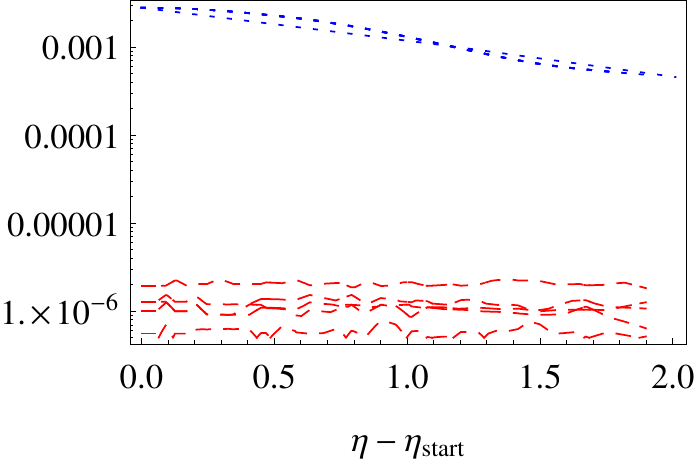}
\caption{Differences $\frac{|\bar{\phi}_{256^3} -\bar{\phi}_{512^3}|}{\bar{\phi}_{512^3}}$ (dashed) and standard deviations $\sigma_{512^3}$ (dotted) for runs on lattices of different sizes. \protect\label{size_dependence}}
\end{minipage}
\end{figure}

For each set of initialisation conditions we run at least one simulation on the lattice of the size of $512^3$ and five simulations on the lattice of the size of $256^3$. 
First five runs on the lattice of the constant size were the base for analysis of statistical fluctuations of results obtained from simulations.
Our code populates randomly the values of the field strength on the lattice at the initialisation, based on the probability distribution \eqref{gauss_distribution}. To be trustworthy our simulation should display only a weak dependence of the evolution of statistical variables on fluctuations of the field strength at the initialisation as far as they correspond to the same probability distribution. The plot in figure \ref{statistical_error} presents variances (dashed lines) 
\begin{equation}
\frac{|\bar{\phi}_i - \left\langle \bar{\phi} \right\rangle|}{\left\langle \bar{\phi} \right\rangle}
\end{equation}
(where average $\langle \cdot \rangle$ is over performed runs and $\bar{\cdot}$ means averaging over the lattice) of the evolution of the mean value of the field strength as a~function of conformal time for five different runs on the lattice of the size $256^3$. 
These variances are compared with the standard deviation (dotted lines) $\sigma_i$ (understood as average over the lattice) of the field strength. We found that the standard deviation of $\bar{\phi}$ is much bigger than differences of the value of $\bar{\phi}$ between different runs.

Secondly comparing results for different lattice sizes we can investigate for numerical artefacts in our implementation. The dependence of results on the size of the lattice should be weak for large enough lattice sizes. The plot in figure \ref{size_dependence} shows the comparison of calculated evolution of the mean value of the field strength $\bar{\phi}$ on two lattice sizes $512^3$ and $256^3$. The differences between results 
\begin{equation}
\frac{|\bar{\phi}_{256^3} -\bar{\phi}_{512^3}|}{\bar{\phi}_{512^3}}
\end{equation}
of five runs on the smaller lattice and one run on the bigger lattice (dashed lines) are much smaller then the standard deviation $\sigma_{512^3}$ (an~average over the lattice) of the field strength computed on the big lattice. 

\section{Algorithm for computing the spectrum of gravitational waves\label{algorithm}}
Computation of the spectrum of GWs based on General Relativity uses the Einstein equation coupled to the eom of the Higgs field. We will use an algorithm based on the linear perturbation expansion of the metric tensor around the Friedman-Robertson-Walker background solution:
\begin{equation}
g = dt^2 - a^2(t) \left(\delta_{ij} + h_{ij}\right) dx^i dx^j = a^2(\eta) \left(d\eta^2 -(\delta_{ij} + h_{ij}) dx^i dx^j\right). 
\end{equation}
GWs are transverse waves, so $h_{ij}$ must be transverse $\partial_i h_{ij}=0$ and traceless $h_{ii}=0$. Linearised Einstein equation leads to equation
\begin{equation}
{\partial_\eta}^2 h_{ij} +2 \frac{\partial_\eta a}{a} \partial_\eta h_{ij} - \Delta h_{ij} = \frac{2}{{M_{Pl}}^2} T^{TT}_{ij}, \label{Einstein}
\end{equation}
where $T^{TT}$ is the transverse-traceless part of the stress tensor $T$ of the Higgs field. In our perturbative expansion the stress tensor $T$ is approximated by
\begin{equation}
T_{ij} = \partial_{x^i} \phi \partial_{x^j} \phi.
\end{equation}
Let us introduce
\begin{equation}
\bar{h} _{ij}(\eta,x):=a(\eta) h_{ij}(\eta,x)
\end{equation}
and its Fourier transform
\begin{equation}
\widehat{\bar{h}}_{ij}(\eta,k)=\int_{\mathbb{R}^3} d^3 x\ e^{-ikx} \bar{h}_{ij}(\eta,x). \label{Fourier}
\end{equation}
Using these variables we can rewrite the Einstein equation \eqref{Einstein} as
\begin{equation}
{\partial_\eta}^2 \widehat{\bar{h}}_{ij} + \left(|k|^2- \frac{{\partial_\eta}^2 a}{a}\right) \widehat{\bar{h}}_{ij} = \frac{2 a}{{M_{Pl}}^2} \widehat{T^{TT}}_{ij}. \label{transformed}
\end{equation}
In the radiation dominated era, which is our primary interest, term $ \frac{1}{a}{\partial_\eta}^2 a$ vanishes \cite{Dufaux:2007pt}. The simplified version of the above equation can be solved using the retarded Green function:
\begin{equation}
\widehat{\bar{h}}_{ij}(\eta,k) = \frac{2}{{M_{Pl}}^2} \int_{\eta_i}^{\eta} d\eta' \frac{\sin \left(|k|\left(\eta - \eta'\right)\right)}{|k|} a(\eta') \widehat{T^{TT}}_{ij} (\eta',k),\label{source_solution}
\end{equation}
where $\eta_i$ is the value of conformal time before which the source $\widehat{T^{TT}}_{ij}$ appeared. The solution for the case of the vanishing source reads
\begin{equation}
\widehat{\bar{h}}_{ij}(\eta,k) = \frac{1}{k} \left(\partial_\eta \widehat{\bar{h}}_{ij}\right)(\eta_f,k) \sin \left(|k|\left(\eta - \eta_f\right)\right) + \widehat{\bar{h}}_{ij}(\eta_f,k) \cos \left(|k|\left(\eta - \eta_f\right)\right), \label{free_solution}
\end{equation}
where $\eta_f$ is the initial conformal time for this solution. 
Assuming that $\eta_i$ is the time of creation of domain walls and $\eta_f$ is the time of their decay we can obtain the perturbation $h_{ij}$ of the metric tensor at the time of matter-radiation equality, by matching solutions of \eqref{source_solution} and \eqref{free_solution} at $\eta_f$.

Another advantage of using the Fourier transform $\widehat{\bar{h}}_{ij}$ is that the traverse-traceless part of energy-momentum tensor can be expressed in compact form
\begin{equation}
\widehat{T^{TT}}_{ij} (k) = \mathcal{O}_{ijlm} (k) \widehat{T}_{ij} (k) = \mathcal{O}_{ijlm} (k) \widehat{\left(\partial_{x^i} \phi \partial_{x^j} \phi\right)} (k), \label{tranverse-traceless}
\end{equation}
using projection operators
\begin{subequations}\label{projection_O}
\begin{equation}
\mathcal{O}_{ijlm} (k) := P_{il} (k) P_{jm} (k) - \frac{1}{2} P_{ij} (k) P_{lm} (k),
\end{equation}
where
\begin{equation}
P_{ij}(k) : = \delta_{ij} - \frac{k_i k_j}{|k|^2}.
\end{equation}
\end{subequations}
According to eq. \eqref{tranverse-traceless} we calculated six independent arrays of $\partial_{x^i} \phi \partial_{x^j} \phi$ using second-order method and then using fast Fourier transform to obtain energy-momentum tensor $\widehat{T^{TT}}_{ij}$. Projection operators $\mathcal{O}$ were calculated in advance for three different families of directions: along edges ($k_i \propto \delta_{1i}, \delta_{2i}, \delta_{3i}$), along diagonals of walls ($k_i \propto \delta_{1i} + \delta_{2i}, \delta_{2i} + \delta_{3i}, \delta_{3i} + \delta_{1i}$) and along the diagonal ($k_i \propto \delta_{1i} + \delta_{2i} + \delta_{3i}$) of the cubic array of momenta.

As indicated in \cite{Dufaux:2007pt} the average energy density $\rho_{gw}$ of gravitational waves with sub-horizon lengths (i.e. ones which can be observed directly) can be approximated as follows:
\begin{equation}
\rho_{gw} = \frac{1}{4} {M_{Pl}}^2 \left\langle \sum_{i,j} \left(\partial_t h_{ij}\right)^2 \right\rangle \approx \frac{{M_{Pl}}^2}{4 a^4} \left\langle \sum_{i,j} \left(\partial_\eta \bar{h}_{ij}\right)^2 \right\rangle.
\end{equation}
According to \cite{Dufaux:2007pt} the average $\langle \cdot \rangle$ should be understood as an integration over volume $V$ of the linear sizes larger then the wave length and integration over time equal to the period $T$. It is worth stressing that integrating over volume $V$ qualifies us to use orthogonality relation for different wave vectors modes. 
These observation justify the validity of the concept of energy density generated by one mode of waves :
\begin{equation}
\varrho_{gw}(\eta,k):= \frac{{M_{Pl}}^2}{32 \pi^3 {a(\eta)}^4 T} \int d \eta \sum_{i,j} \left|\partial_\eta \widehat{\bar{h}}_{ij}(\eta,k)\right|^2. \label{one_mode_density} 
\end{equation}
After substituting solution from eq. \eqref{free_solution} into eq. \eqref{one_mode_density} one obtains:
\begin{equation}
\varrho_{gw}(\eta,k) = \frac{{M_{Pl}}^2}{32 \pi^3 {a(\eta)}^4 V} \sum_{i,j} \left(\left|\left(\partial_\eta \widehat{\bar{h}}_{ij}\right)(\eta_f,k)\right|^2 + |k|^2 \left|\widehat{\bar{h}}_{ij}(\eta_f,k)\right|^2\right).
\end{equation}
Energy density generated by one mode $\varrho_{gw}(\eta,k)$ can be expressed as:
\begin{equation}
\begin{split}
\varrho_{gw}(\eta,k) =& \frac{1}{16 \pi^3 {M_{Pl}}^2 {a(\eta)}^4 V} \sum_{i,j} \left[\left|\int_{\eta_i}^{\eta_f} d\eta' \cos \left(|k|\left(\eta - \eta'\right)\right) a(\eta') \widehat{T^{TT}}_{ij} (\eta',k)\right|^2 \right.\\
&\left. + \left| \int_{\eta_i}^{\eta_f} d\eta' \sin \left(|k|\left(\eta - \eta'\right)\right) a(\eta') \widehat{T^{TT}}_{ij} (\eta',k)\right|^2\right],
\end{split}\label{density_final}
\end{equation}
using the mentioned matching and the solution given by eq. \eqref{source_solution}.

In this paper we are mostly interested in the spectrum of gravitational waves' energy density per unit logarithmic frequency interval:
\begin{equation}
\frac{d \rho_{gw}}{d \log |k|} (\eta,k) = |k|^3 \int_{S^2} d \Omega_k \varrho_{gw}(\eta,k) =:\frac{1}{(2 \pi)^2 {M_{Pl}}^2 {a(\eta)}^4 V} S(\eta,k), \label{energy_density}
\end{equation}
where $\int_{S^2} d \Omega_k$ denotes the integration over the direction of the wave vector $k$. Last equality in eq. \eqref{energy_density} defines the function $S$ which was calculated in our numerical simulations.\footnote{Let us stress that $S$ is the different function then the function $S_k$ defined in \cite{Dufaux:2007pt}.} Using the solution given by eq. \eqref{density_final} the function $S$ can be expressed as:
\begin{equation}
\begin{split}
S(\eta,k) =& \frac{|k|^3}{4 \pi} \int_{S^2} d \Omega_k \sum_{i,j,l,m} \left[\left|\int_{\eta_i}^{\eta_f} d\eta' \cos \left(|k|\left(\eta - \eta'\right)\right) a(\eta') \mathcal{O}_{ijlm} \widehat{T}_{lm} (\eta',k)\right|^2 \right.\\
&\left. + \left| \int_{\eta_i}^{\eta_f} d\eta' \sin \left(|k|\left(\eta - \eta'\right)\right) a(\eta') \mathcal{O}_{ijlm} \widehat{T}_{lm} (\eta',k)\right|^2\right].
\end{split}\label{S_expression}
\end{equation}
In cubic lattice simulations the integral over directions of wave vector proves difficult, because only small number of vectors on the lattice have the same length. For example, there are only six different vectors of the length equal to $1$ (in units of the lattice spacing): $(1,0,0)$, $(0,1,0)$, $(0,0,1)$ and $(-1,0,0)$, $(0,-1,0)$, $(0,0,-1)$ belonging to three different edges (coinciding with three axes). In the isotropic Universe, the spectrum of gravitational waves generated from the network of many domain walls should be nearly isotropic, so we approximated the function $S$ using the average over vectors from three families: along edges, along diagonals of walls and along the diagonal of the cubic array of momenta.
\subsection{Spectrum today}
During the evolution of the Universe gravitational waves produced by the network of domain walls were stretched by the expansion. In this subsection we will present methods used to estimate the red-shifted spectrum. As noted previously the spectrum before the epoch of the matter-radiation equality can be computed using solution \eqref{free_solution} from eq. \eqref{energy_density}:
\begin{equation}
\frac{d \rho_{gw}}{d \log |k|} (\eta_{EQ},k) = \frac{{a(\eta_{dec})}^4}{{a(\eta_{EQ})}^4} \frac{2 \pi |k|^3 }{{M_{Pl}}^2 {a(\eta_{dec})}^4} S(\eta_{dec},k) = \frac{{a(\eta_{dec})}^4}{{a(\eta_{EQ})}^4} \frac{d \rho_{gw}}{d \log |k|} (\eta_{dec},k),
\end{equation}
where $\eta_{EQ}$ is the conformal time at which the energy densities of matter and radiation are equal and $\eta_{dec}$ is conformal time at which the decay of domain walls has ended. The factor $\frac{{a(\eta_{dec})}^4}{{a(\eta_{EQ})}^4}$ which is determined by the evolution of the scale factor in the radiation domination epoch (eq. \eqref{scale_factor}), can be easily computed.

The value of the Hubble constant at the time of equality of matter and radiation energy densities can be calculated assuming simple scaling of these densities:
\begin{equation}
{H_{EQ}}^2=\frac{2 {H_0}^2 {\Omega_M}^4}{{\Omega_R}^3}=2\times10^{-55} \frac{\textrm{eV}^2}{\hbar^2}, \label{equlibrium}
\end{equation}
from present values of the Hubble constant $H_0$ and fractions of the critical density $\Omega_M$, $\Omega_R$. The value of the Hubble constant at conformal time after the decay of domain walls can be computed from parameters of the simulation:
\begin{equation}
H_{dec}=\frac{1}{a_{dec}\eta_{dec}} \left(1-\frac{a_{in}}{a_{dec}}\right) =10^{19} \left(1-\frac{a_{in}}{a_{dec}}\right) \left(\frac{10^{-10}\ \hbar\textrm{GeV}^{-1}}{a_{dec} \eta_{dec}}\right) \frac{\textrm{eV}}{\hbar}. \label{simulation}
\end{equation}
The ratio $\frac{a_{in}}{a_{dec}}$ is relevant for the theories in which radiation domination epoch begins at very low energy scales. In most inflationary scenarios this value is negligible.
Using \eqref{equlibrium} and \eqref{simulation} we obtain
\begin{equation}
\frac{{a(\eta_{dec})}}{{a(\eta_{EQ})}} = \sqrt{\frac{{H_{EQ}}}{{H_{dec}}}}= 7.1\times10^{-24} \left(\frac{10^{19} \frac{\textrm{eV}}{\hbar}}{H_{dec}}\right)^{\frac{1}{2}}.
\end{equation}
Furthermore in order to obtain the present spectrum of GWs we need to calculate how the energy density of GWs has changed from the epoch of equality to the present day. Assuming that the energy density of GWs scales as $a^{-4}$ we can write:
\begin{equation}
\frac{d \rho_{gw}}{d \log |k|} (\eta_{0},k) = (1+z_{EQ})^{-4} \frac{{a(\eta_{dec})}^4}{{a(\eta_{EQ})}^4} \frac{d \rho_{gw}}{d \log |k|} (\eta_{dec},k),
\end{equation}
where $\eta_{0}$ is the present time and $z_{EQ}$ is the red-shift to the epoch of matter-radiation equality. The energy density of gravitational waves $\frac{d \rho_{gw}}{d \log |k|} (\eta)$ is usually presented as a fraction $\Omega_{gw} (\eta)$ of the critical density $\rho_{cr}(\eta):= {M_{Pl}}^2 H^2(\eta)$.

Finally we must calculate present frequencies associated with our wave vectors. The wavelength $\lambda_{dec}$ of the GW with the comoving wave vector $k$ at the time of the decay of domain walls satisfies:
\begin{equation}
k a(\eta_{dec}) \lambda_{dec}=2 \pi. \label{wavelength_dec}
\end{equation}
On the other hand the present wavelength $\lambda_{0}$ satisfies similar equation:
\begin{equation}
k a(\eta_0) \lambda_0=2 \pi. \label{wavelength_0}
\end{equation}
Equating $\frac{k}{2\pi}$ from eqs. \eqref{wavelength_dec} and \eqref{wavelength_0} we estimated the red-shift of the wave frequency to be equal to:
\begin{equation}
f_{0}=\frac{a(\eta_{dec})}{a(\eta_0)} \frac{k}{2 \pi}=5.07\times 10^{6} \left(\frac{10^{19} \frac{\textrm{eV}}{\hbar}}{H_{dec}}\right)^{\frac{1}{2}} \left(\frac{k}{10^{10}\ \frac{\textrm{GeV}}{\hbar\textrm{c}}}\right)\ \textrm{Hz}. \label{redshited_f}
\end{equation}
\section{Spectrum of gravitational waves\label{spectrum}}
Previous studies of GWs \cite{Hiramatsu:2013qaa} emitted during the decay of domain walls have shown that existence of the maximum at the frequency corresponding to the Hubble radius ${H_{dec}}^{-1}$ at the time of the decay is a generic property of their spectrum. Using eq. \eqref{redshited_f} one can estimate that the spectrum of GWs from SM domain walls is peaked at the value of the order of 
\begin{equation}
f_{peak} \approx 3.18 \times 10^{7} \left(\frac{H_{dec}}{10^{19} \frac{\textrm{eV}}{\hbar}}\right)^{\frac{1}{2}} \ \textrm{Hz}, \label{approx_peak_frequency}
\end{equation}
where $\eta_{dec}$ is the conformal time at which our domain walls have decayed. Using eq. \eqref{simulation} and our observation that the decay usually ends at the conformal time of the order of \mbox{$10^{-9}\ \hbar\textrm{GeV}^{-1}$} one estimates
\begin{equation}
f_{peak} \sim 7.96\times 10^{6} \ \textrm{Hz}.
\end{equation}
For nearly degenerate minima one can estimate the energy density of GWs at the peak using semi-analytic expression \cite{Hiramatsu:2013qaa,Kitajima:2015nla} 
\begin{equation}
\left. \Omega_{gw} (\eta_{dec}) \right|_{peak} = \left. \frac{1}{\rho_c(\eta_{dec})} \left(\frac{d \log \rho_{gw}}{d \log \eta}\right) \right|_{peak} 
= \frac{\tilde{\epsilon}_{gw} \mathcal{A}^2 {\sigma_{wall}}^2}{24 \pi {H_{dec}}^2 {M_{Pl}}^4},
\end{equation}
where $\sigma_{wall}$ is the surface energy density (tension) of domain walls and $\tilde{\epsilon}_{gw}$, $\mathcal{A}$ are respectively efficiency and scaling parameters determined in numerical simulations of $\lambda \phi^4$ model \cite{Kitajima:2015nla} to be equal to $\tilde{\epsilon}_{gw} \simeq 0.7$ and $\mathcal{A} \simeq 0.8$. The resulting tension of SM domain walls is $1.16686 \times 10^{29}\ \frac{\textrm{GeV}^3}{(c \hbar)^2}$. Finally one estimates:
\begin{equation}
\left. \Omega_{gw} (\eta_{dec}) \right|_{peak} \sim 2 \times 10^{-38} \left(\frac{10^{19} \frac{ \textrm{eV}}{\hbar}}{H_{dec}}\right)^{2}. \label{approx_omega}
\end{equation}
The relic density of GWs will further decreases after the emission and is estimated to be of the order of
\begin{equation}
\left. \Omega_{gw} (\eta_0) h^2 \right|_{peak} \sim 9 \times 10^{-42} \left(\frac{10^{19} \frac{ \textrm{eV}}{\hbar}}{H_{dec}}\right)^{2} \label{approx_omega_0}
\end{equation}
today. 

The estimated peak frequency is higher than frequencies measurable in planned detectors of GWs and the estimated relic density is many orders of magnitude lower then predicted sensitivities for these detectors. However according to eqs. \eqref{approx_peak_frequency} and \eqref{approx_omega} decreasing the value of $H_{dec}$ will simultaneously decrease the peak frequency $f_{peak}$ and increase the relic density $\Omega_{gw}$. 

Calculating the spectrum of gravitational waves in lattice simulations encounters even more complications. The modified eom \eqref{PRS} with $\alpha=3$ and $\beta=0$ cannot be used in this calculation, because the modification of eom disturbs the dynamics of the short wavelength fluctuations. For the unmodified eom the width of the domain walls decreases as $\propto a^{-2}$. The width of the domain walls expressed in units of lattice spacing cannot be too small in order to model correctly the profile of walls. This requirement significantly restricts the dynamical range of the simulation and only late domain walls ($\eta_{start} = \mathcal{O}(10^{-9}\ \textrm{GeV}^{-1})$) can be investigated in our simulation. Moreover as noticed in \cite{Hiramatsu:2013qaa}, algorithm presented in section \ref{algorithm} produces a spectrum that diverges as $k^3$ (due to the factor $|k|^3$ in \eqref{S_expression}) for random initialisation of the field strength. Following \cite{Hiramatsu:2013qaa} we introduced a cut-off scale of the order of the width of domain walls in the Fourier transform of the initialisation distribution. 

\begin{figure}
\subfloat[]{\label{eta=10}
\includegraphics[width=0.47\textwidth]{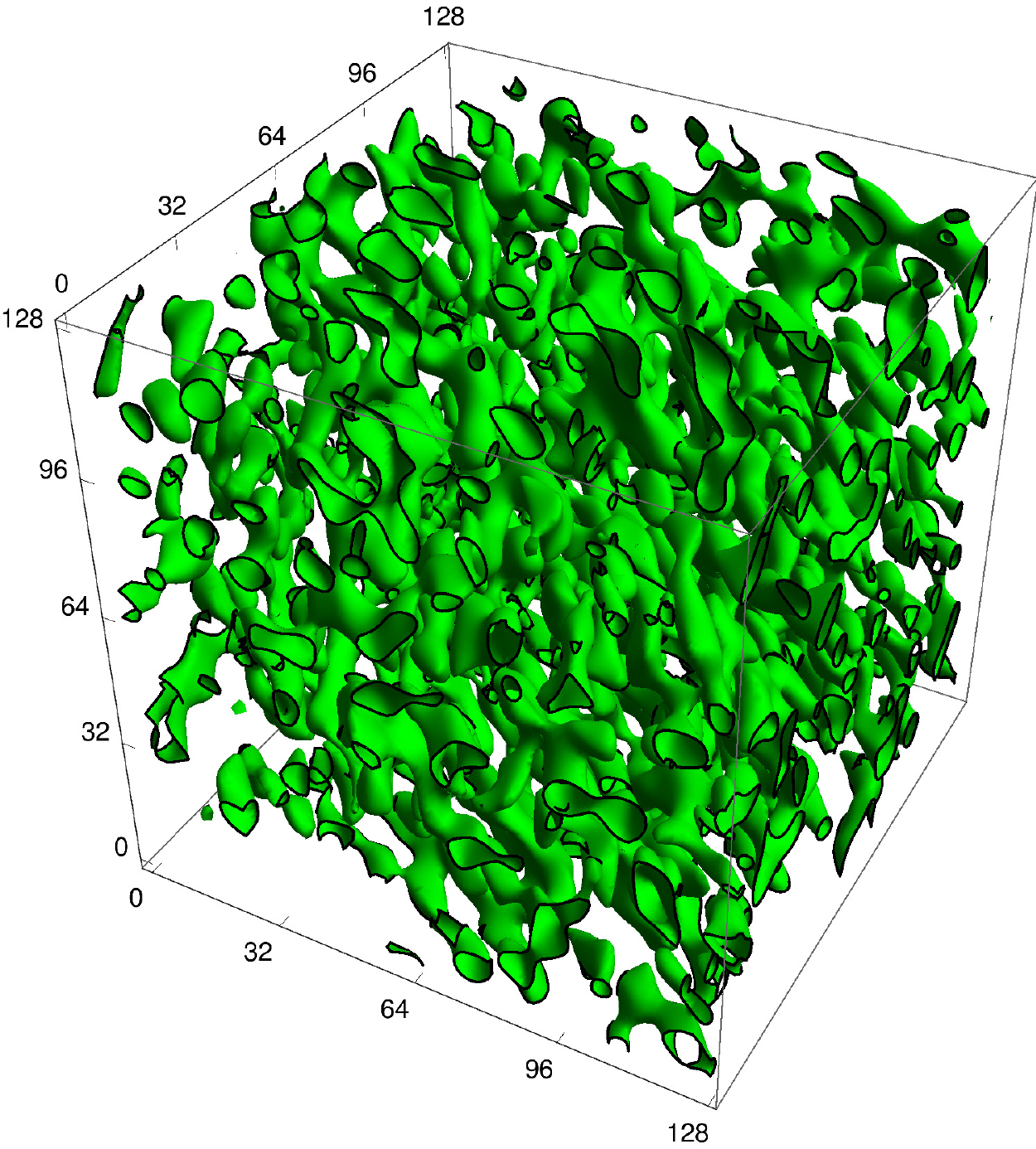}
}
\subfloat[]{\label{eta=12}
\includegraphics[width=0.47\textwidth]{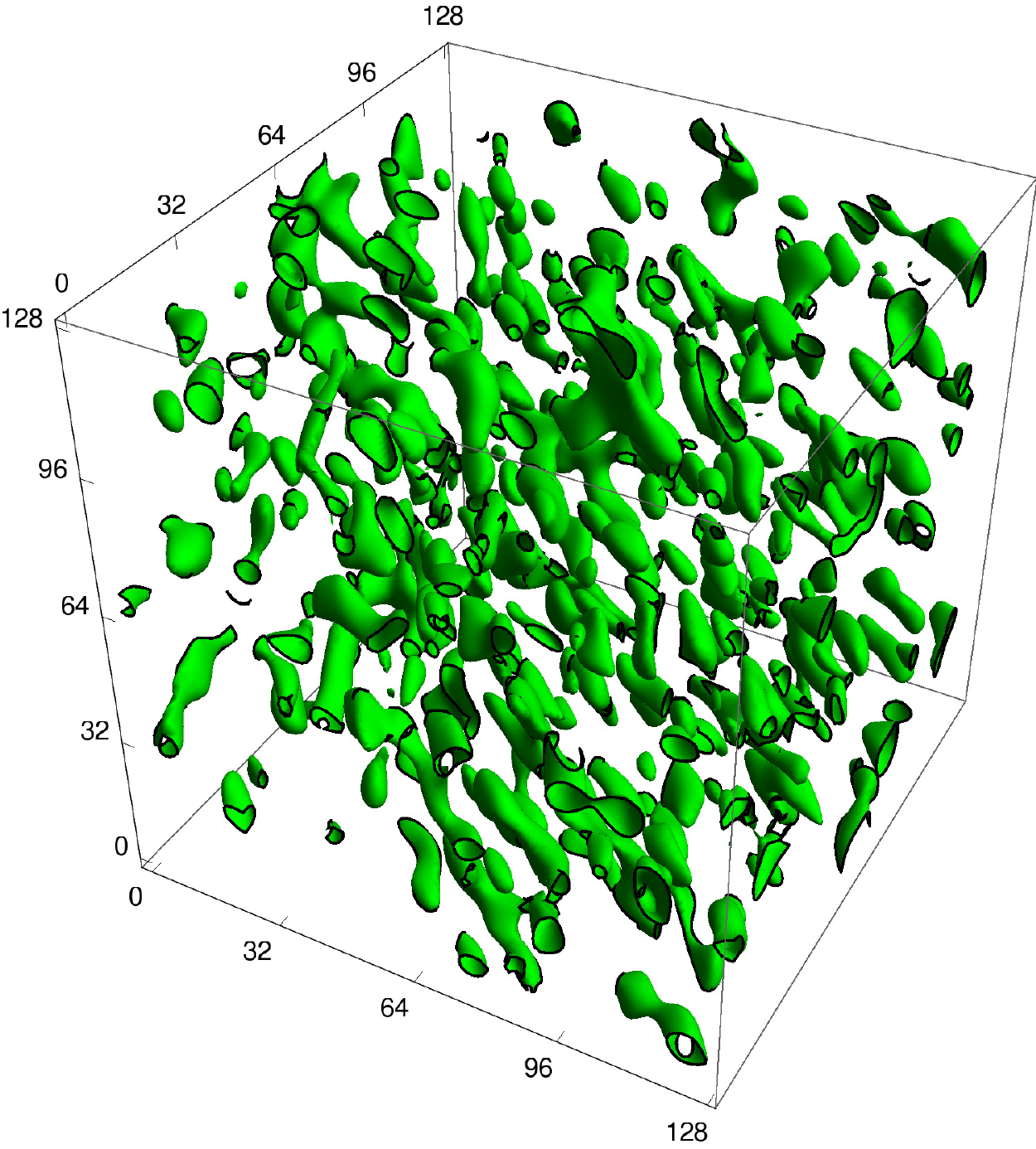}
}\\
\subfloat[]{\label{eta=13}
\includegraphics[width=0.47\textwidth]{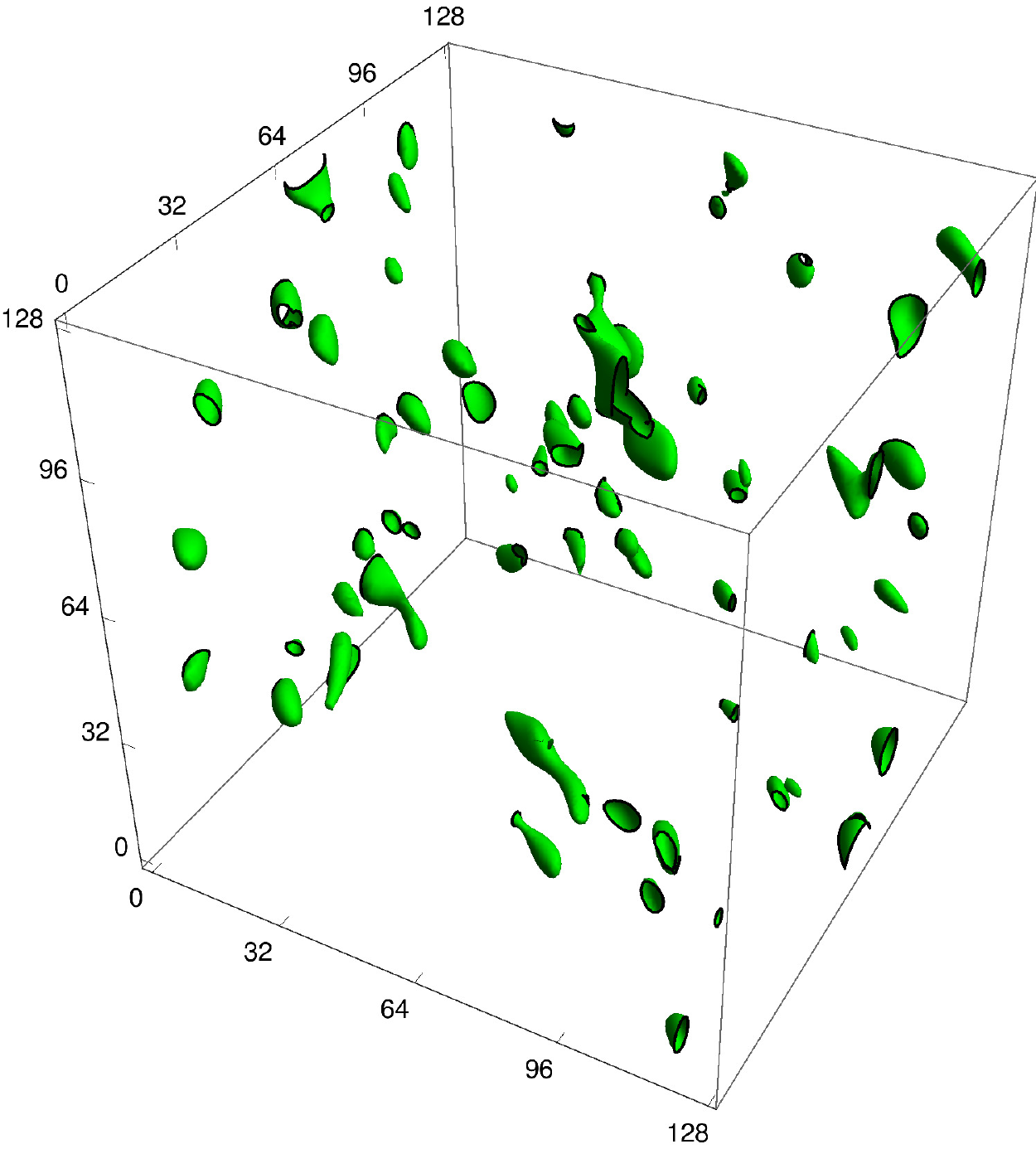}
}
\subfloat[]{\label{eta=14}
\includegraphics[width=0.47\textwidth]{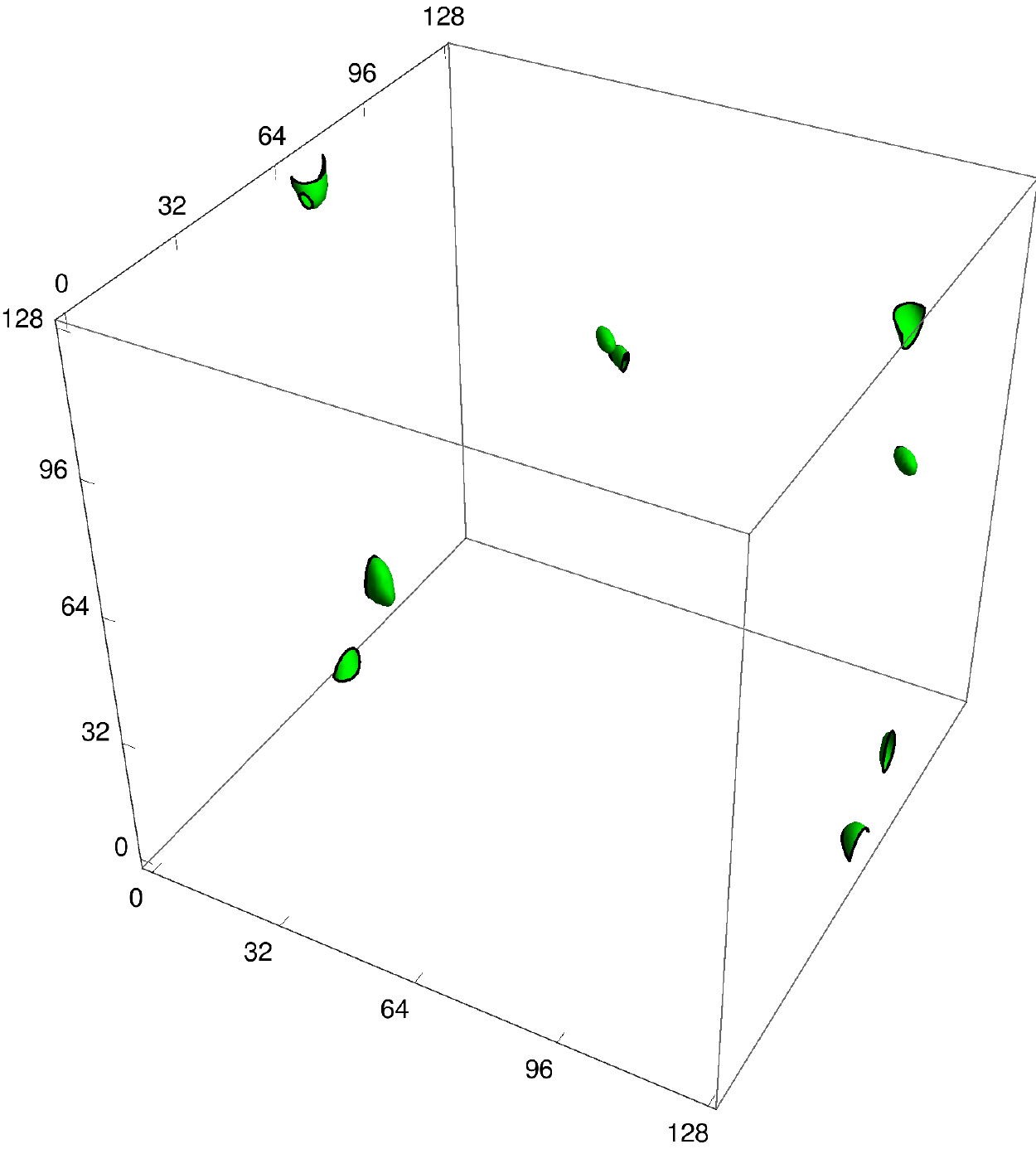}
}
\caption{Visualization of the isosurface of the field strength $\phi$ corresponding the value $v_{max}$ at four different conformal times: $\eta=10^{-9}\ \textrm{GeV}^{-1}$ \protect\subref{eta=10} and  $\eta=1.2 \times 10^{-9}\ \textrm{GeV}^{-1}$ \protect\subref{eta=12}, $\eta=1.3 \times 10^{-9}\ \textrm{GeV}^{-1}$ \protect\subref{eta=13}, $\eta=1.4 \times 10^{-9}\ \textrm{GeV}^{-1}$\protect\subref{eta=14}. Lengths are given in units of the lattice spacing i.e. $10^{-10}\ \textrm{GeV}^{-1}$. \protect\label{network}}
\end{figure}

The evolution of the network of domain walls with the setup used for calculation of the spectrum of GWs (however performed on the smaller lattice size $128^3$) is presented in the figure \ref{network}. The isosurface of the field strength $\phi$ corresponding to the position of the maximum of the effective potential $v_{max}$ approximated by the interpolation from values at lattice sites is visualized for four different conformal times: $\eta=10^{-9}\ \textrm{GeV}^{-1}$ (initialization time) and  $\eta=1.2 \times 10^{-9}\ \textrm{GeV}^{-1}$, $\eta=1.3 \times 10^{-9}\ \textrm{GeV}^{-1}$, $\eta=1.4 \times 10^{-9}\ \textrm{GeV}^{-1}$. In the first panel \ref{eta=10} we see a~rich structure of the network produced by the initialization algorithm which decays during the simulation (panels \ref{eta=12} and \ref{eta=13}), ending with only a~few bubbles (panel \ref{eta=14}).

\begin{figure}
\begin{minipage}{0.48\textwidth}
\includegraphics[width=\textwidth]{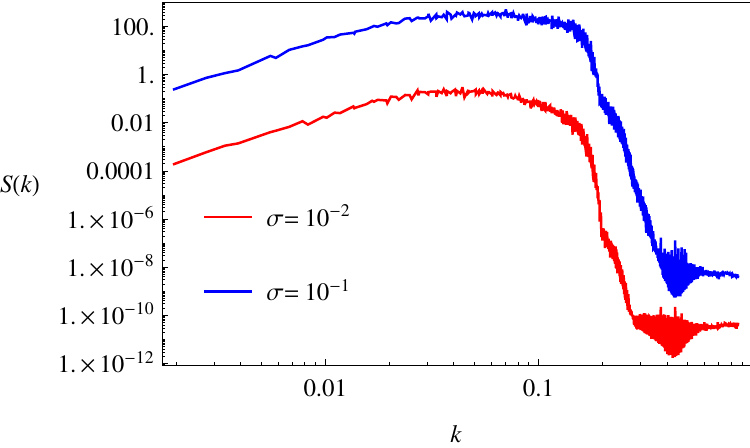}
\caption{The function $S(k)$ \eqref{S_expression} computed in numerical simulation (giving the spectrum of GWs after rescaling) expressed in units of lattice spacing. \protect\label{S_function}}
\end{minipage}%
\hspace{0.5cm}
\begin{minipage}{0.48\textwidth}
\includegraphics[width=\textwidth]{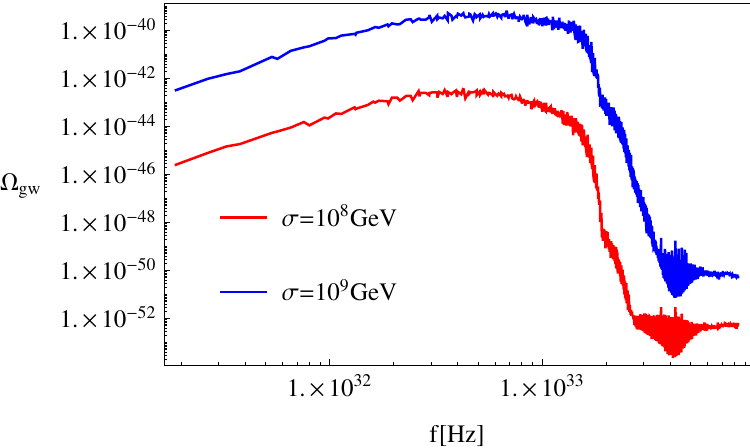}
\caption{Spectrum of gravitational waves $\Omega_{gw}$ emitted from SM domain walls at the time of the decay. \protect\label{rho_GW}}
\end{minipage}
\end{figure}

Figure \ref{S_function} presents the function $S(k)$ defined by eq. \eqref{S_expression} expressed in units of the lattice spacing. $S(k)$ was computed as an~average over $5$ simulations which were initialised with condition $\theta=v_{max}+\sigma$ for $\sigma = 10^{8}\ \textrm{GeV}$ and $\sigma = 10^{9}\ \textrm{GeV}$ at the initial conformal time $\eta_{start}=10\ l$ with $a_{start}:=a(\eta_{start}) = \frac{1}{5}$. Simulations ended at the conformal time equal to $\eta_{end}=20\ l$ with $a_{end}:=a(\eta_{end}) = 4$. The width of domain walls at this time was of the order of $10\ l$. The obtained spectrum is peaked at wave vectors of the order of $k_{peak} \sim 0.05\ l^{-1}$, corresponding to the frequency \mbox{$f_{peak} \sim 3\times10^{8}\ \textrm{GeV}$}. We assumed that $a_{in}=0$ and the value of Hubble constant is equal to $\frac{1}{a_{end}\eta_{end}}$. In figure \ref{rho_GW} we plotted the spectrum of GWs $\Omega_{gw}$ at the end of the simulation as a~function of the frequency $f$ in units of $\textrm{Hz}$.

\begin{figure}
\subfloat[]{\label{a0=0}
\includegraphics[width=0.49\textwidth]{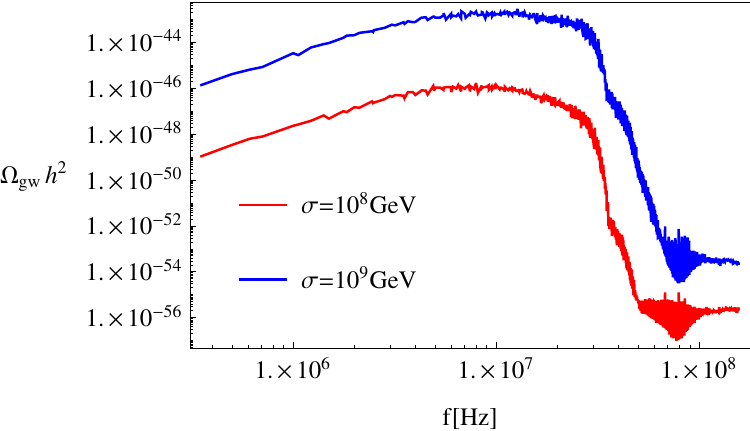}
}
\subfloat[]{\label{a0=0.9}
\includegraphics[width=0.49\textwidth]{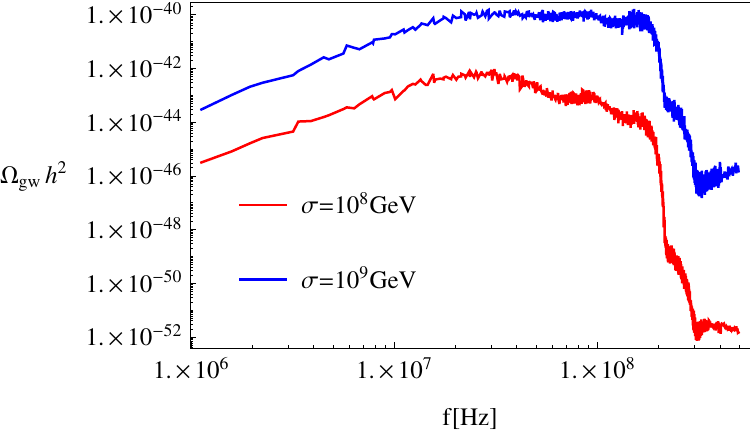}
}
\caption{Present spectrum of gravitational waves $\Omega_{gw}$ emitted from SM domain walls for $\frac{a_{in}}{a_{dec}}=0$ ast in standard cosmology \protect\subref{a0=0} and $\frac{a_{in}}{a_{dec}}=0.1$ \protect\subref{a0=0.9}. \protect\label{Omega_GW}}
\end{figure}

In figure \ref{a0=0} we plotted the spectrum of GWs $\Omega_{gw}$ at present time and in figure \ref{a0=0.9} the present spectrum for the case of non-standard cosmology $\frac{a_{in}}{a_{end}}=0.1$. 

Our simulations predict the amplitude lower than the one predicted by the semi-analytical method. However in our simulations we observe a dependence of the amplitude of the spectrum on the amplitude of initial fluctuations. Extrapolating our results we find that for fluctuations of the order of $10^{10}\ \textrm{GeV}$, the amplitude of the spectrum predicted by semi-analytical method can be reproduced. Moreover the position of the peak in the spectrum calculated in simulations scales differently with $H_{dec}$ than it is predicted by the semi-analytical method. The time of the decay of domain walls is weakly dependent on the value of $H_{dec}$. However, the red-shift is lower for lower values of $H_{dec}$ and the frequency of the peak grows. The discrepancy arises, because the semi-analytical method assumes emergence of scaling regime in the evolution of domain walls (period in the evolution with nearly constant number of domain walls in comoving horizon) which is absent in the evolution of investigated domain walls. If domain walls decay through the scaling regime then most of the energy is carried away by GWs emitted at the end of the decay (this is the least red-shifted contribution) and these GWs build up the peak of the spectrum. As numerical simulations have shown, Higgs domain walls decayed shortly after they entered the comoving horizon, and the value of the Hubble constant $H_{dec}$ determines (in our parametrization) only the moment in the history of the Universe at which the decay happened. Although the redshift of the peak of the spectrum of Higgs domain walls is given by the value of $H_{dec}$, GWs corresponding to the peak were not necessarily emitted at the time corresponding to the Hubble constant equal to $H_{dec}$.

\section{Summary\label{conclusions}}
In this paper we investigated the possibility of formation of a~network of SM domain walls.
 We concentrated on evolution and decay of such a network and left open the question of an underlying cosmological model which leads to initial conditions facilitating its creation. We described these processes using numerical simulations performed on the lattice using PRS algorithm. Main observables in our simulations were the conformal time at which the network of domain walls decays and the spectrum of GWs emitted during this decay. 

In section \ref{width} we show how the width of domain walls in a simple $\lambda \phi^4$ model can be obtained with the first integral. We use this method to estimate the width of SM domain walls. This approximation leads to the width of the order of \mbox{$4\times 10^{-9}\ \frac{\hbar c}{\textrm{GeV}}$}. The width of domain walls in our simulations must be a~few times bigger than the lattice spacing. We use lattice spacing equal to \mbox{$l=10^{-10}\ \frac{c \hbar}{\textrm{GeV}}$} which results in units for (conformal) time \mbox{$10^{-10}\ \frac{\hbar}{\textrm{GeV}}$} and energy $10^{10}\ \textrm{GeV}$.

Our simulations show that domain walls, interpolating between minima of the RG improved effective potential of the SM, decay shortly after their formation. Decay times range from \mbox{$8 \times 10^{-11}\ \hbar\textrm{GeV}^{-1}$} to \mbox{$3 \times 10^{-9}\ \hbar\textrm{GeV}^{-1}$}. Such short lifetimes exclude a~scenario in which SM domain walls dominate the Universe leading to large distortion of the CMBR. On the other hand networks of domain walls could produce large fluctuations in the energy density of matter if the walls survive till late times or become re-created. These fluctuations could be a~seed for structure formation and would leave their imprints in the CMBR. Investigation of such fluctuations  is beyond the scope of this paper and we postpone it for future work. It is interesting to note, that the creation of the network which ends up in the electroweak vacuum percolating through the Universe is not as difficult to obtain as one may think, although it requires certain tuning of initial conditions.

The second signature of the existence of the network of SM domain walls are gravitational waves emitted during the decay of domain walls. GWs have an advantage that they interact with matter  very weakly after their emission and the spectrum of their energy density has stayed nearly unchanged during the evolution of the Universe except for the red-shift.

\begin{figure}[t]
\includegraphics[width=\textwidth]{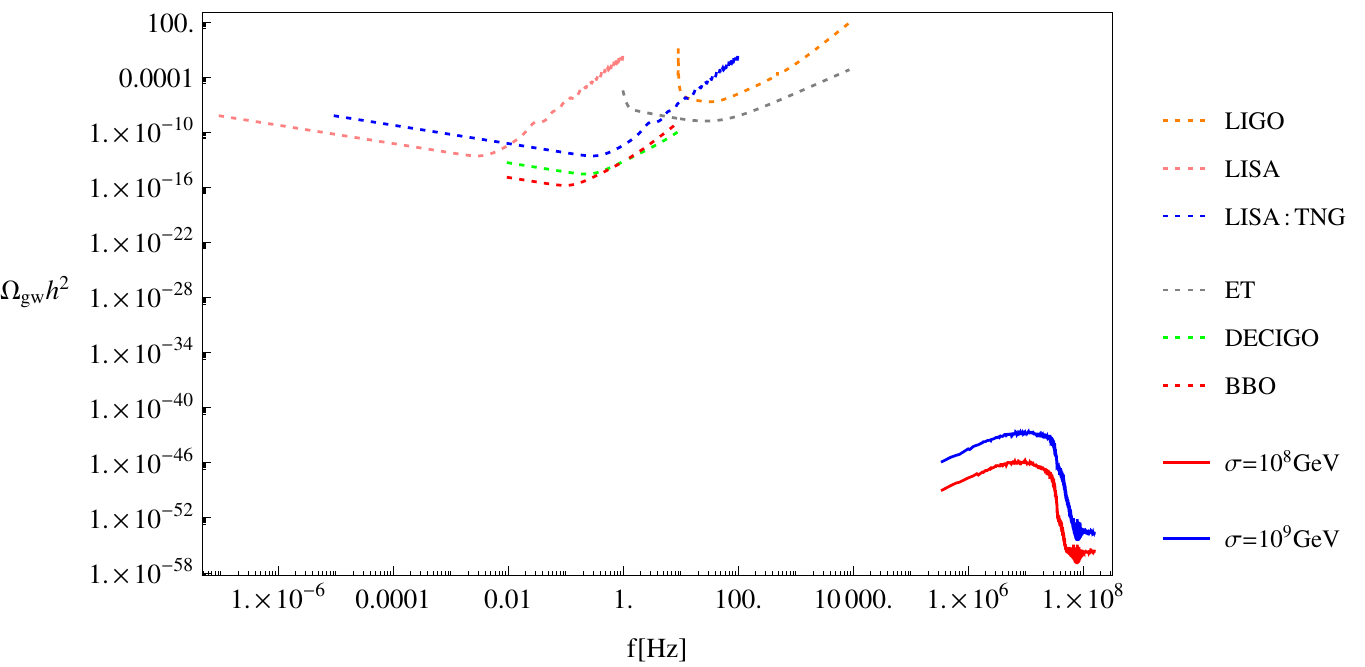}
\caption{Predicted sensitivities (dashed) for future GWs detectors: aLIGO, ET, LISA, LISA:TNG, DECIGO and BBO compared with the spectrum of GWs (solid) calculated in lattice simulations for the initial values of $\sigma=10^8,\ 10^9\ \textrm{GeV}$ and the standard cosmology. \label{sensitivity}}
\end{figure}

We calculate the spectrum of GWs in our numerical simulation using the algorithm (described in section \ref{algorithm}) which was previously used to predict the spectrum of primordial GWs produced during inflation and GWs produced by domain walls in $\lambda \phi^4$ model. In figure \ref{sensitivity} obtained spectra (solid) of GWs are compared with predicted sensitivity (dashed) of future detectors: aLIGO \cite{TheLIGOScientific:2014jea}, ET \cite{Hild:2010id}, LISA, LISA:TNG \cite{Larson:1999we}, DECIGO and BBO \cite{Yagi:2011wg}. 

Short inspection of the plot from figure \ref{sensitivity} reveals that assumption about the standard cosmological evolution and validity of SM up to very high scales exclude the detection of GWs emitted by SM domain walls in upcoming years. However, both of these assumptions can be weakened. The present energy density of GWs produced by domain walls could be greater if the evolution of the Universe before formation of domain walls was different then in the standard scenario. Models predicting very low scale of inflation \cite{Artymowski:2016ikw} or including new components of energy density which shorten the radiation domination period \cite{Lewicki:2016efe,Huang:2016odd} would result in larger $\frac{a_{in}}{a_{dec}}$ (or lower $H_{dec}$). However this possibility is hard to investigate using lattice simulations due to their small dynamical range and require semi-analytical extrapolation. Moreover recent experimental bounds on the size of nonrenormalizable interactions of SM particles are very weak. The recent studies \cite{Lalak:2014qua} have shown that the inclusion of the nonrenormalizable operators in the Lagrangian density (which corresponds to the assumption that the scale at which the SM breaks down is smaller than the Planck scale $M_{Pl}$) modify the behaviour of the effective potential $V_{\text{eff}}$ at high scales and can even make EW vacuum stable. The scenario of nearly degenerate minima of Higgs potential due to inclusion of $|H|^6$ operator has been recently investigated in \cite{Kitajima:2015nla}. The analysis presented in \cite{Kitajima:2015nla} shows that for small difference in potential density of two minima, the decay time of domain walls is longer, leading to the increased energy density of emitted GWs. Moreover for very small differences the energy density can be high enough for GWs to be detectable in future interferometer detectors.
The future detection of GWs emitted from decaying Higgs domain walls, although not likely, would point toward non-standard cosmological scenarios and its significance for both particle physics and cosmology is hard to overestimate. 

\acknowledgments{
This work has been supported by the Polish NCN grants DEC-2012/04/A/ST2/00099 and 2014/13/N/ST2/02712, ML was also supported by the doctoral scholarship number \\ 2015/16/T/ST2/00527.
ZL thanks DESY Theory Group for hospitality. This work was supported by the German Science Foundation (DFG) within the Collaborative Research Center (SFB) 676 "Particles, Strings and the Early Universe".
}

\appendix
\section{Proof of lack of soliton solutions for potential with non-degenerate minima}\label{app:nosoliton}
\begin{figure}[t]
\centering
\includegraphics[width=0.5\textwidth]{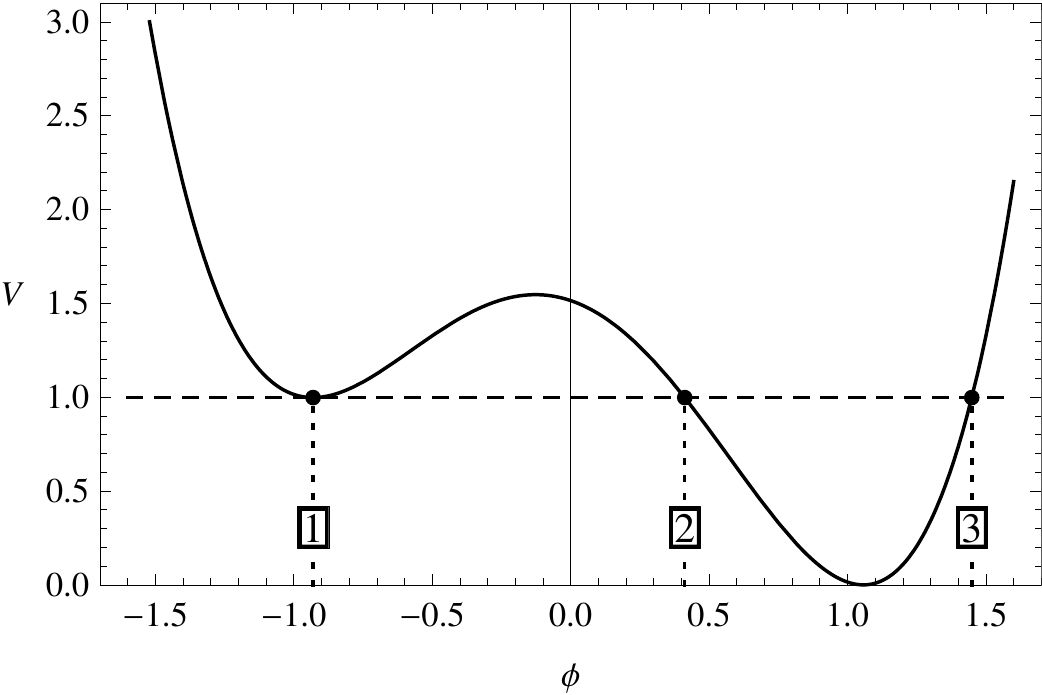}
\caption{The qualitative behaviour of the generic potential of the form \eqref{toy_lagrangian_density}. The arguments with the same value of the potential as the minimum with higher energy (horizontal dashed line) are denoted by $1$, $2$ and $3$ ($1$ is the mentioned minimum). \label{3_points}}
\end{figure}

We will show that time-independent soliton solution for planar domain wall does not exist if minima of a~potential are non-degenerate. We can see this using eq. \eqref{energy}. Let us fix the value of the potential at the stable minimum $v_0$ to be equal to $0$ and at the unstable minimum $v_{\delta V}$ as $\delta V$. If we assume that the solution is going towards the deeper minimum to the left (analogously as the solution \eqref{soliton}) i.e.
\begin{equation}
\lim_{x\to - \infty} \varphi(x) = v_0,
\end{equation}
then the constant $E$ \eqref{energy} is equal to $0$ for this solution ($\varphi_{E=0}$). The solution $\varphi_{E=0}$ satisfies the equation:
\begin{equation}
\left( \varphi'_{E=0} \right)^2 = 2 V\left(\varphi_{E=0}\right).
\end{equation}
It is now obvious that $ \varphi'_{E=0}$ can be equal to $0$ only in the $v_0$ minimum, so the only bounded solution $\varphi_{E=0}$ is the trivial one $\varphi_{E=0}=v_0$. In the case when
\begin{equation}
\lim_{x\to - \infty} \varphi(x) = v_{\delta V},
\end{equation}
the constant $E$ \eqref{energy} will take the value $\delta V$. There are at least three points $\phi_{\delta V}$ which satisfy
\begin{equation}
V\left(\phi_{\delta V}\right)=\delta V,
\end{equation}
and $ \varphi'_{E=\delta V}(x_0)$ can be equal to $0$ for the $x_0$ such that $\varphi_{E=\delta V}(x)=\phi_{\delta V}$. The potential from eq. \eqref{toy_lagrangian_density} plotted in figure \ref{3_points} illustrates the generic situation.
The three mentioned points are denoted as $1$, $2$ and $3$, where point $1$ corresponds to the local minimum with energy $\delta V>0$. There is a solution for which 
\begin{equation}
\lim_{x\to - \infty} \varphi(x) = v_{\delta V},
\end{equation}
however as the field $\varphi$ reaches point $2$ the first derivative $\varphi'_{E=\delta V}$ goes to $0$. Using \eqref{toy_eom}:
\begin{equation}
 - \varphi''_{E=\delta V} = - \frac{\partial V}{\partial \phi} \left(\varphi_{E=\delta V}\right),
\end{equation}
one finds that the second derivative $ \varphi''_{E=\delta V}$ is negative at $2$, so after reaching this value, the value of the field will decrease and finally
\begin{equation}
\lim_{x\to + \infty} \varphi(x) = v_{\delta V}.
\end{equation}
\section{Discretisation of equations of motion}
In recent paper we used a~discretisation scheme at first proposed in \cite{Press:1989yh}. It was widely used in the past for numerical simulations of the dynamics of domain walls for example in \cite{Lalak:2007rs,Lazanu:2015fua}. Let us consider the following generalisation of eq. \eqref{SM_eom}
\begin{equation}
\frac{\partial^2 \phi}{\partial \eta^2} + \alpha \left(\frac{d \log{a}}{d \log{\eta}}\right) \frac{1}{\eta} \frac{\partial \phi}{\partial \eta} - \Delta \phi + a^\beta \frac{\partial V}{\partial \phi}=0, \label{PRS}
\end{equation}
proposed in \cite{Press:1989yh}. The equation \eqref{PRS} with values $\alpha=3$ and $\beta=0$ combined with used discretisation are known as PRS algorithm. 

We used the Crank-Nicholson scheme for space-like derivatives which gives us the second order expression for the Laplacian of the field $\phi$:
\begin{equation}
\left(\Delta \phi\right)^n_{i,j,k} = \phi^n_{i+1,j,k}+\phi^n_{i-1,j,k}+\phi^n_{i,j+1,k}+\phi^n_{i,j-1,k} +\phi^n_{i,j,k+1}+\phi^n_{i,j,k-1}-6 \phi^n_{i,j,k}, \label{laplacian_discretisation}
\end{equation}
where $\phi^n_{i,j,k}$ denotes the field value at the lattice point with coordinates $i$, $j$, $k$ and $n$ numbers integration steps. The "staggered leapfrog" was used for the second-order time derivatives. Discretised equations are as follows:
\begin{subequations}\label{discretisation}
\begin{align}
\ddot{\phi}^{n}_{i,j,k} =& \frac{-2\delta\dot{\phi}^{n-\frac{1}{2}}_{i,j,k}+\Delta\eta\left[\left(\Delta \phi\right)^n_{i,j,k}-a^\beta \frac{\partial V}{\partial \phi}(\phi^n_{i,j,k})\right]}{(1+\delta)},\\ \label{eom_discretisation}
\dot{\phi}^{n+\frac{1}{2}}_{i,j,k} =& \dot{\phi}^{n-\frac{1}{2}}_{i,j,k}-\frac{\Delta\eta}{(1+\delta)}\ddot{\phi}^{n}_{i,j,k},\\
\phi^{n+1}_{i,j,k} =& \phi^n_{i,j,k}+\Delta\eta^n \dot{\phi}^{n+\frac{1}{2}}_{i,j,k},
\intertext{where we defined:}
\delta :=& \frac{1}{2} \alpha \frac{\Delta\eta}{\eta} \frac{d \log{a}}{d \log{\eta}}.
\end{align}
\end{subequations}

This discretisation scheme (in fact any discretisation based on fixed lattice spacing), even through successfully used for modeling of dynamics of domain walls in the past, poses certain problems in the case of the RG improved SM potential $V_{\text{SM}}$. For very high field strengths $|h| \gg v_{max}$ the derivative of the potential $\frac{\partial V_{\text{SM}}}{\partial |h|}$ can be qualitatively approximated as $\frac{\partial V_{\text{SM}}}{\partial |h|} \propto - |h|^3$. On the other hand our discretisation of the Laplacian given by eq. \eqref{laplacian_discretisation} scales with the central value $\phi^n_{i,j,k}$ (with values at neighboring points fixed) linearly. Thus for high enough field strengths $\phi$ the potential term in the discretisation of the eom \eqref{eom_discretisation} will dominate $|a^\beta \frac{\partial V}{\partial \phi}(\phi^n_{i,j,k})| \gg |\left(\Delta \phi\right)^n_{i,j,k}|$. However as we saw in section \ref{width} the formation of domain walls corresponds to a~situation when these two terms are of the same order. We conclude that for field strengths high enough to $|a^\beta \frac{\partial V}{\partial \phi}(\phi^n_{i,j,k})| \gg |\left(\Delta \phi\right)^n_{i,j,k}|$ our simulation must break down. Another reasoning which exposes the same issue is as follows: introducing field theory on the lattice corresponds to setting an~UV cutoff scale (equal to the inverse of the lattice spacing) on momenta, however increasing the field strength is equivalent to increasing the energy scale over the introduced UV cutoff. As a result we are not allowed to treat consistently initial field configurations with extremely high field strengths using the constant lattice spacing approach. However, the reported simulations were performed well within the limits of applicability of our numerical technology. 

\section{Optimisation of time step}
The adaptive time step has proved to be very useful in our simulations. Using this method decreases the number of needed time steps leading to reduction of both: a~computational time and errors coming from finite accuracy of the representation of floating point numbers. As stressed in section \ref{SM_potential} the effective potential of the SM is much more computationally demanding than previously investigated ones. We developed more advanced methods for time step optimisation than the ones that were proposed in \cite{Lalak:2007rs}.

In our simulations we have used the discretisation schemes which are second order accurate both in time and space. Using Taylor expansion of the exact (hypothetical) solution $\Phi$ of eq. \eqref{SM_eom} one can check that eqs. \eqref{discretisation} give the approximation of $\Phi$ which is accurate up to terms which are proportional to 
\begin{equation}
\left(\Phi(\eta+\Delta\eta)-\Phi(\eta)\right)-\left(\phi^{n+1} - \phi^n\right)\propto\frac{\partial^3 \Phi}{\partial \eta^3}(\eta) {(\Delta\eta)}^3 + \BigO{{\Delta\eta^4}},
\end{equation}
where $\frac{\partial^3 \Phi}{\partial \eta^3}$ can be estimated as follows:
\begin{equation}
\frac{\partial^3 \Phi}{\partial \eta^3}(\eta) \approx \frac{\ddot{\phi}^{n} - \ddot{\phi}^{n-1}}{\Delta\eta}.
\end{equation}
The relative error generated in integration step can be estimated from:
\begin{equation}
\frac{\left(\Phi(\eta+\Delta\eta)-\Phi(\eta)\right)-\left(\phi^{n+1} - \phi^n\right)}{\phi^n} \propto \frac{\ddot{\phi}^{n} - \ddot{\phi}^{n-1}}{\phi^n{\Delta\eta}^2}. \label{relative_error}
\end{equation}
In our simulation the time step $\Delta\eta$ is calculated from condition that the maximal value (over the lattice) of the relative error estimated according to eq. \eqref{relative_error} must be smaller than some small constant number $\kappa$. 

\bibliographystyle{JHEP}
\bibliography{GWDWSM}
\end{document}